\documentclass[journal,onecolumn,12 pt,times]{IEEEtran}
\usepackage[margin=1 in]{geometry}
\usepackage{lineno,hyperref}
\usepackage{slashbox,pict2e}
\usepackage{xspace}
\usepackage{amsmath,amssymb,array}
\usepackage{booktabs}
\usepackage{listings}
\usepackage[T1]{fontenc}
\usepackage{textcomp}
\usepackage[sort&compress,square,comma,authoryear]{natbib}
\usepackage{graphicx}
\usepackage[svgnames]{xcolor}
\usepackage{caption,subcaption}%
\usepackage{setspace}
\usepackage{natbib}

\graphicspath{{Figures/}}

\newcommand{\ExaGeoStatR}	{{\em {ExaGeoStatR}}\xspace}
\newcommand{\ExaGeoStat}		{{\em {ExaGeoStat}}\xspace}
\newcommand{\exageostatr}		{{\em {ExaGeoStatR}}\xspace}

\newcommand{\hicma}				{{\em {HiCMA}}\xspace}
\newcommand{\dplasma}				{{\em {DPLASMA}}\xspace}
\newcommand{\starsh}				{{\em {STARS-H}}\xspace}
\newcommand{\chameleon}		{{\em {Chameleon}}\xspace}
\newcommand{\starpu}				{{\em {StarPU}}\xspace}
\newcommand{\hwloc}				{{\em {hwloc}}\xspace}
\newcommand{\nlopt}					{{\em {NLopt}}\xspace}
\newcommand{\parsec}					{{\em {PaRSEC}}\xspace}
\newcommand{\gsl}						{{\em {GSL}}\xspace}

\newcommand{\blas}					{{\em {BLAS}}\xspace}
\newcommand{\R}					{{\em {R}}\xspace}
\newcommand{\geor}					{{\em {GeoR}}\xspace}
\newcommand{\fields}					{{\em {fields}}\xspace}
\newcommand{\inla}					{{\em {INLA}}\xspace}
\newcommand{\randomfields}					{{\em {RandomFields}}\xspace}
\newcommand{\biggp}					{{\em {bigGP}}\xspace}
\newcommand{\spbayes}					{{\em {spBayes}}\xspace}
\newcommand{\mkl}					{{\em {Intel MKL}}\xspace}
\newcommand{\openblas}					{{\em {OpenBLAS}}\xspace}
\newcommand{\openmp}					{{\em {OpenMP}}\xspace}

\newcommand{\C}{{\em C}\xspace}
\newcommand{\CUDA}{{\em CUDA}\xspace}
\newcommand{\MPI}{{\em MPI}\xspace}
\modulolinenumbers[5]

\usepackage{amssymb}
\usepackage{pifont}
\newcommand{\xmark}{\color{red}\ding{55}}%
\newcommand{\cmark}{\color{green}\ding{51}}%

\ifCLASSINFOpdf
\else
\fi
\begin{document}
%
\title{Large-scale Environmental Data Science with ExaGeoStatR}
%
%
%


\author{
\begin{spacing}{1.3}
\normalsize{
    \IEEEauthorblockN{Sameh Abdulah, Yuxiao Li, Jian Cao, Hatem Ltaief, David E. Keyes, \\Marc G. Genton, and Ying Sun} 
    
    \IEEEauthorblockA{{Computer, Electrical and Mathematical Sciences and Engineering Division,\\
King Abdullah University of Science and Technology (KAUST),
Thuwal, Saudi Arabia}
}
}
\end{spacing}
}

\maketitle
\begin{spacing}{1.2}

\begin{abstract}

Parallel computing in exact Gaussian process calculations becomes necessary for avoiding computational and memory restrictions associated with large-scale environmental data science applications. The exact evaluation of the Gaussian log-likelihood function requires $O(n^2)$ storage and $O(n^3)$ operations where $n$ is the number of geographical locations. Thus, exactly computing the log-likelihood function with a large number of locations requires exploiting the power of existing parallel computing hardware systems, such as shared-memory, possibly equipped with GPUs, and distributed-memory systems, to solve this exact computational complexity. In this paper, we present \ExaGeoStatR, a package for exascale Geostatistics in \R that supports a parallel computation of the exact maximum likelihood function on a wide variety of parallel architectures. Furthermore, the package allows scaling existing Gaussian process methods to a large spatial/temporal domain. Prohibitive exact solutions for large geostatistical problems become possible with \ExaGeoStatR. Parallelization in \ExaGeoStatR depends on breaking down the numerical linear algebra operations in the log-likelihood function into a set of tasks and rendering them for a task-based programming model. The package can be used directly through the \R environment on parallel systems without the user needing any \C, \CUDA, or \MPI knowledge. Currently, \ExaGeoStatR supports several maximum likelihood computation variants such as exact, Diagonal Super Tile (DST) and Tile Low-Rank (TLR) approximations, and Mixed-Precision (MP). \ExaGeoStatR also provides a tool to simulate large-scale synthetic datasets. These datasets can help assess different implementations of the maximum log-likelihood approximation methods. Herein, we show the implementation details of \ExaGeoStatR, analyze its performance on various parallel architectures, and assess its accuracy using synthetic datasets with up to 250K observations. The experimental analysis covers the exact computation of \ExaGeoStatR to demonstrate the parallel capabilities of the package. We provide a hands-on tutorial to analyze a sea surface temperature real dataset. The performance evaluation involves comparisons with the popular packages \geor, \fields, and \biggp for exact Gaussian likelihood evaluation. The approximation methods in \ExaGeoStatR are not considered in this paper since they were analyzed in previous studies.

\end{abstract}

\end{spacing}


\begin{spacing}{1.8}

%
\IEEEpeerreviewmaketitle

\section{Introduction}
\label{sec:intro}
Massive data require a delicate analysis to extract meaningful information
from their complex structure. With the availability of large data volumes
coming from different sources, an objective of data science 
is to combine a set of principles and techniques from data mining, 
machine learning, and statistics to support a better understanding of the 
given data. Like other sciences, environmental science has
been very much impacted by the field of data science. Indeed, a vast pool of 
techniques has been developed to understand spatial and spatio-temporal data 
coming from intelligent sensors or satellite images. However, new 
challenges have arisen with today's data sizes that require a unique treatment paradigm.

Environmental applications deal with measurements regularly or irregularly located across a geographical region. Gaussian processes (GPs), or Gaussian Random Fields (GRFs), are among the most useful tools in  applications for fitting spatial datasets. For the last two decades, GPs have been used extensively to model spatial data and are able to cover a wide range of spatial data with different specifications~\citep{gelfand2016}.
Spatial data can also be described as random effects, hence they can often be modeled with a Gaussian distribution. Therefore, the dependencies in spatial data can be described through a Gaussian process.

GRFs also serve as building 
blocks for numerous non-Gaussian models in spatial statistics such as trans-Gaussian 
random fields, mixture GRFs and skewed GRFs. For example, \cite{doi:10.1080/01621459.2016.1205501} introduced the Tukey $g$-and-$h$ random field which modifies the GRF by a flexible 
form of variable transformation; \cite{andrews1974scale}, \cite{west1987scale}, and \cite{rue2005gaussian} considered the scale-mixture of Gaussian distributions 
and GRFs; and \cite{allard2007new} and \cite{azzalini2005skew} proposed many 
skewed GRFs and their variations. However, likelihood-based inference methods 
for GP models are computationally expensive for large-scale spatial datasets. It is crucial 
to provide fast computational tools to fit a GP model to large-scale spatial datasets often 
available in many real-world applications.

Specifically, suppose $Z({\mathbf s})$ is a stationary GRF with mean function $m({\mathbf s})$ and covariance function $C({\mathbf s},{\mathbf s}^\prime)$, and we observe data on a domain $D\subset \mathbb{R}^d$ at $n$ locations, $\mathbf {s}_1,\ldots, \mathbf {s}_n$. Then, the random vector $\{Z(\mathbf {s}_1),\ldots, Z(\mathbf{s}_n)\}^\top$ is assumed to follow a multivariate Gaussian distribution:
\begin{equation}
 \forall \{\mathbf{s}_1, \ldots, \mathbf{s}_n\} \subset D, \quad
 \{Z(\mathbf{s}_1), \ldots, Z(\mathbf{s}_n) \}^\top \sim {\cal N}_n(\boldsymbol{\mu},{\mathbf{\Sigma}}),
\label{eq:distribution}
\end{equation}
where $\boldsymbol{\mu}=\{m(\mathbf{s}_1),\ldots,m(\mathbf{s}_n)\}^\top$ and ${\mathbf{\Sigma}}$ are the mean vector and the covariance matrix of the $n$-dimensional multivariate normal distribution. Given $\boldsymbol{\mu}$ and ${\mathbf{\Sigma}}$, the (Gaussian) likelihood of observing $\mathbf{z} = \{z(\mathbf{s}_1),  \ldots, z(\mathbf{s}_n) \}^{\top}$ at the $n$ locations is
\begin{equation}
L(\boldsymbol{\mu},\mathbf{\Sigma})=\frac{1}{(2\pi)^{n/2}|{\mathbf{\Sigma}}|^{1/2}}\exp\left\{ -\frac{1}{2}(\mathbf{z} - \boldsymbol{\mu})^{\top}{\mathbf{\Sigma}}^{-1}(\mathbf{z}-\boldsymbol{\mu}) \right\}.
\label{eq:likelihood}
\end{equation}

The $(i,j)$-th element of $\mathbf{\Sigma}$ is $\Sigma_{ij}=C(\mathbf{s}_i,\mathbf{s}_j)$, where the covariance function $C(\mathbf{s}_i,\mathbf{s}_j)$ is assumed to have a parametric form with unknown vector of parameters $\boldsymbol \theta$. Various classes of covariance functions can be found in e.g. \cite{cressie}. For simplicity, in this work, we assume the mean vector $\boldsymbol{\mu}$ to be zero to focus on estimating the covariance parameters. We choose the most popular isotropic Mat{\'e}rn covariance kernel, which is specified as,
\begin{equation}
\Sigma_{ij}=C(\|\mathbf{s}_i-\mathbf{s}_j\|)=\frac{\sigma^2}{2^{\nu-1}\Gamma(\nu)}\left(\frac{\|\mathbf{s}_i-\mathbf{s}_j\|}{\beta}\right)^{\nu}\mathcal{K}_{\nu}\left(\frac{\|\mathbf{s}_i-\mathbf{s}_j\|}{\beta}\right),
\label{eq:matern}
\end{equation}
where $\|\mathbf{s}_i-\mathbf{s}_j\|$ is the Euclidean  
distance between $\mathbf{s}_i$ and $\mathbf{s}_j$, $\Gamma(\cdot)$ is the gamma function,  $\mathcal{K}_{\nu}(\cdot)$ is the modified Bessel function of the second kind of order $\nu$, and $\sigma^2$, $\beta>0$, and $\nu>0$ are the key parameters of the covariance function controlling the variance, spatial range, and smoothness, respectively. The Mat{\'e}rn covariance kernel is highly flexible and includes the exponential ($\nu=1/2$)  and Gaussian ($\nu=\infty$) kernels as special cases. The variance, spatial range, and smoothness parameters, $\sigma^2$, $\beta$, and $\nu$ determine the properties of the GRF.

The typical inference for GRFs includes parameter estimation, stochastic simulation, and kriging (spatial prediction). Among these tasks, parameter estimation, or model fitting, is the most time-consuming. Once the parameters are estimated, one can easily simulate multiple realizations of the GRF or obtain predictions at new locations. To obtain the maximum likelihood estimator (MLE), we need to optimize the likelihood function in Equation \eqref{eq:likelihood} over $\boldsymbol{\theta}=(\sigma^2, \beta, \nu)^{\top}$. However, the likelihood for a given $\boldsymbol{\theta}$ requires computing the inverse ($\mathbf{\Sigma(\boldsymbol{\theta})}^{-1}$) and the determinant ($|\mathbf{\Sigma(\boldsymbol{\theta})}|$) of the covariance matrix $\mathbf{\Sigma}(\boldsymbol{\theta})$, and performing the triangular linear solve ($\mathbf{\Sigma(\boldsymbol{\theta})}^{-1}\mathbf{Z}$).
The most time-consuming operation to compute the likelihood function is the Cholesky factorization for $\mathbf{\Sigma}(\boldsymbol{\theta})$, which requires $O(n^3)$ operations and $O(n^2)$ memory. This factorization is required to compute the inverse and the determinant of $\mathbf{\Sigma}(\boldsymbol{\theta})$. Due to the likelihood estimation complexity, the standard methods and traditional algorithms for GRFs are computationally infeasible for large datasets.

On the other hand, technological advances in sensor networks along with the investments in data monitoring, collection, resource management provide massive open-access spatial datasets \citep{Finley:2015aa}. The unprecedented data availability and the challenging computational complexity call for novel methods, algorithms, and software packages to deal with modern ``Big Data'' problems in spatial statistics.

A broad literature focuses on developing efficient methodologies by approximating the covariance function in the GP model, so that the resulting covariance matrix is easier to compute. \cite{sun2012geostatistics}, \cite{bradley2016comparison}, and \cite{liu2018gaussian} systematically reviewed these methods. Some popular approximation methods are covariance tapering \citep{furrer2006covariance, kaufman2008covariance}, discrete process convolutions \citep{higdon2002space, lemos2009spatio}, fixed rank kriging \citep{cressie2008fixed}, lattice kriging \citep{nychka2015multiresolution}, and predictive processes \citep{banerjee2008gaussian, finley2009improving}. Meanwhile, some studies proposed to approximate the Gaussian likelihood function using conditional distributions \citep{vecchia1988estimation,KG2021} or composite likelihoods \citep{varin2011overview,eidsvik2014estimation}, and some seek for equivalent representation of GPs using spectral density \citep{fuentes2007approximate} and stochastic partial differential equations \citep{lindgren2011explicit}.
 
A recent direction of this research aims at developing parallel algorithms \citep{bigGP, katzfuss2017parallel, datta2016hierarchical, guhaniyogi2018meta} and using modern computational architectures, such as multicore systems, GPUs, and supercomputers, in order to avoid insufficient approximation of the GP  \citep{simpson2012think,stein2014limitations}. Aggregating computing power through High-Performance Computing (HPC) becomes an important tool in scaling existing software in different disciplines to handle the exponential growth of datasets generated in these fields~\citep{vetter2013contemporary}. 
However, the literature lacks a well-developed HPC software that practitioners can use to support their applications with HPC capabilities. Although most studies provide reproducible source codes, they are difficult to extend to new applications, especially when the algorithms require certain hardware setups. {\em R} \citep{ihaka1996r} is the most popular software in statistics, applied analytics, and interactive exploration of data by far. As a high-level language, however, {\em R} is relatively weak for high-performance computing compared to lower-level languages, such as {\em C}, {\em C++}, and {\em Fortran}. Scaling statistical software and bridging high-performance computing with the \R language can be performed using two different strategies. One strategy has been followed by the {\em pbdR} \citep{pbdR2012} project, Programming with Big Data in \R, which transfers the HPC libraries to the \R environment by providing a high-level  \R interface to a set of HPC libraries such as {\em MPI}, {\em ScaLAPACK}, {\em ZeroMQ}, to name a few. However, one drawback of this strategy is that the \R developer should have enough background in HPC  to be able to use the provided interfaces to scale his/her code. Another strategy that we adopt in this paper is to implement the statistical functions using an HPC-friendly language such as {\em C}. Then it is easier to directly wrap the {\em C} functions into \R functions. In this case, these functions can directly be used inside the \R environment without the need to understand the underlying HPC architectures or the development environment.

This paper presents \ExaGeoStatR, a high-performance package in \R for
large-scale environmental data science and geostatistical applications that depends on a unified  {\em C}-based software called \ExaGeoStat \citep{abdulah2018exageostat}.  \ExaGeoStat is able to fit Gaussian process models and provide spatial predictions and simulations for geostatistics applications in large-scale domains. {\em ExaGeoStat} provides both exact and approximate computations for large-scale spatial datasets. Besides the exact method, the software also supports three approximation methods: Diagonal Super Tile (DST), Tile Low-Rank (TLR), and Mixed-Precision (MP) computations. This study highlights the capabilities of the \ExaGeoStatR exact computations since it can be considered a benchmark for the performance of other computation methods. Moreover, the evaluation of the DST and the TLR approximations has already been covered in \cite{abdulah2018exageostat, abdulah2018parallel, abdulah2019geostatistical,hong2021efficiency,abdulah2021accelerating}. The software also includes a synthetic dataset generator for generating large spatial datasets with the exact prespecified covariance function. Such large datasets can be used to perform broader scientific experiments related to large-scale computational geostatistics applications. Besides its ability to deal with different hardware architectures such as multicore systems, GPUs, and distributed systems, \ExaGeoStatR utilizes the underlying hardware architectures to its full extent. Existing assessments on \ExaGeoStat show the ability of the software to handle up to 3.4 million spatial locations on manycore systems with exact calculations \citep{abdulah2021accelerating}.


Existing \R packages for fitting GRFs include \geor \citep{geoR}, \fields \citep{fields}, \spbayes\citep{Finley:2007aa, Finley:2015aa}, \randomfields \citep{RandomFields1, RandomFields2}, \inla \citep{rue2009approximate, martins2013bayesian}, \biggp \citep{bigGP}. These packages feature different degrees of flexibility as well as computational capacity. The \spbayes package fits GP models in the context of Bayesian or hierarchical modeling based on MCMC. The {\randomfields} package implements the Cholesky factorization method, the circulant embedding method \citep{dietrich1996fast}, and an extended version of Matheron's turning bands method \citep{matheron1973intrinsic} for the maximum likelihood estimation of GRFs. The {\inla} package uses an integrated nested Laplace approximation to tackle additive models with a latent GRF, which outperforms the MCMC method. The {\biggp} package utilizes distributed memory systems through RMPI \citep{gropp1999using} to implement the estimation, prediction, and simulation of GRFs. The packages {\geor} and {\fields} both estimate the GRF covariance structures designed for spatial statistics while \geor provides more flexibilities, such as estimating the mean structure and the variable transformation parameters. Among these popular packages, only the {\biggp} package, according to our knowledge, focuses on distributed computing, which is essential for solving problems in data-rich environments in exact mode. The {\biggp} package was built using the {\em RMPI} and {\em OpenMP} libraries to facilitate GP calculations on manycore systems. The {\biggp} package relies on block-based algorithms to perform the underlying linear algebra operations required to perform the GP calculations. Compared to the {\biggp} package, \exageostatr provide better performance since it depends on the state-of-the-art parallel linear algebra operations through applying tile-based algorithms to perform the required linear solvers.
Table~\ref{tab:pkgs} provides a summary of some of the existing packages for fitting GRFs. The comparison includes common features for each package and if it supports parallel execution or not.

Our package \ExaGeoStatR, at the current stage, performs data generation, parameter estimation, and spatial prediction for the univariate GRF with mean zero and a Mat\'ern covariance structure, which is a fundamental model in spatial statistics. We feature breakthroughs in the optimization routine for the maximum likelihood estimation and the utilization of heterogeneous computational units. Specifically, we build on the optimization library \nlopt \citep{johnson2014nlopt} and provide a unified Application Programming Interface (API) for multicore systems, GPUs, clusters, and supercomputers. The package also supports using the great circle distance on the sphere in constructing the covariance matrix. These parallelization features largely reduce the time-per-iteration in computing exact maximum likelihood estimations compared with existing packages and make GRFs even with $10^6$ locations estimable on hardware accessible to most institutions.

The remainder of this paper is organized as follows. Section~\ref{section2} states the basics of \ExaGeoStatR, and the package components for keen readers. Section~\ref{section3} compares the estimation accuracy and time of \ExaGeoStatR with two aforementioned exact computation packages, i.e., {\geor} and {\fields}, with simulated data. It demonstrates the efficiency that can be gained when the \ExaGeoStatR package utilizes powerful architectures including GPUs and distributed memory systems. Section~\ref{section4} is a tutorial to fit a Gaussian random field to a sea surface temperature dataset with more than ten thousand spatial locations per day and perform kriging with the estimated parameter values. Section~\ref{section5} concludes with the contributions of the \ExaGeoStatR package. In the Appendix, we provide a user guide for installing the \ExaGeoStatR package on different hardware environments.



\section{Software Overview}
\label{section2}

\subsection{ExaGeoStat Outline}

\ExaGeoStat \footnote{https://github.com/ecrc/exageostat} is
a {\em C}-based software that targets environmental
applications through a high-performance parallel implementation of
the Maximum Likelihood Estimation (MLE) operation that is widely
used for geospatial statistics \citep{abdulah2018exageostat}. This software
provides a novel solution to deal with the scaling limitation impact of the
MLE operation by exploiting the computational power of emerging hardware
architectures. \ExaGeoStat permits exploring the MLE computational limits
using state-of-the-art high-performance dense linear algebra libraries by
leveraging a single source code to run on various cutting-edge parallel
architectures with the aide of runtime systems software.
The ``separation of concerns'' philosophy adopted by \ExaGeoStat from the beginning permits improving the software from different perspectives by allowing fast linear solvers and porting the code to different hardware architectures.

\ExaGeoStat 
is developed to solve the MLE problem for a given
set of data observed at  $n$ geographical locations on a large scale and to provide
a prediction solution for unobserved values at new locations. The software
also allows for exact synthetic data generations with a given covariance
function, which can be used to test and compare different approximation methods.
To sum up, \ExaGeoStat includes three primary tools: large-scale synthetic data
generator, the Gaussian maximum likelihood estimator, and the geospatial predictor.

\begin{figure}[t!]
  \centering
  \begin{subfigure}[b]{0.24\linewidth}
    \centering\includegraphics[width=0.7\textwidth, height=3.5cm]{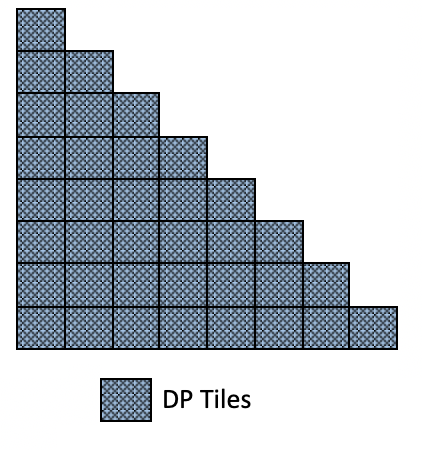}
    \caption{\label{fig:fig1} Fully dense (Exact)}
  \end{subfigure}%
  \begin{subfigure}[b]{0.24\linewidth}
    \centering\includegraphics[width=0.7\textwidth, height=3.5cm]{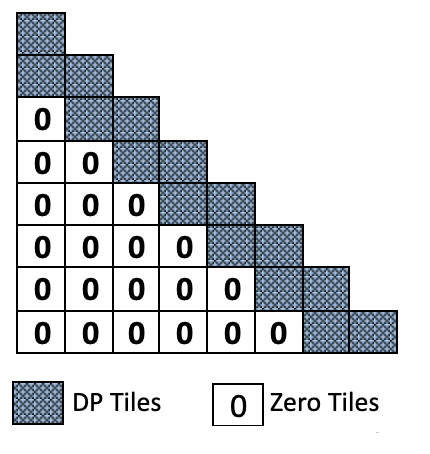}
    \caption{\label{fig:fig2}Diag. Super-Tile (DST)}
  \end{subfigure}
  \begin{subfigure}[b]{0.24\linewidth}
    \centering\includegraphics[width=0.72\textwidth, height=3.7cm]{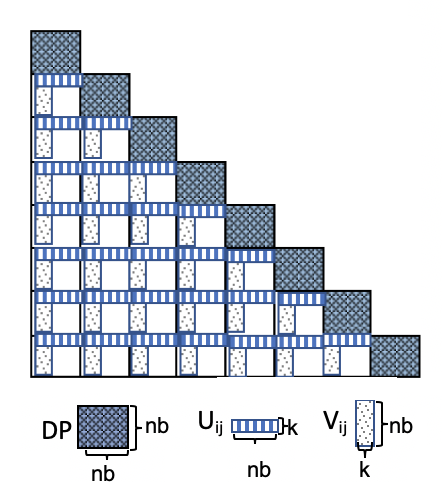}
    \caption{\label{fig:fig3}Tile Low-Rank (TLR)}
      \end{subfigure}
      \begin{subfigure}[b]{0.24\linewidth}
    \centering\includegraphics[width=0.8\textwidth, height=3.6cm]{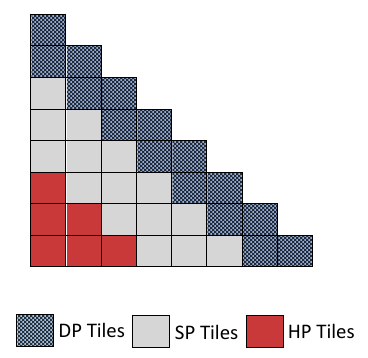}
    \caption{\label{fig:fig4}Mixed-Precision (MP)}
      \end{subfigure}
    
  \caption{ \label{fig:computation_variants} \ExaGeoStat supports various computation methods. The tile is a subset of the matrix with size $ts \times ts$. Here $ts$ should be tuned for performance on parallel systems. DP refers to double-precision (64-bit) tile, SP refers to single-precision (32-bit) tile, and HP refers to half-precision (16-bit) tile.}
\end{figure}

In \ExaGeoStat, the MLE operation is implemented in four different
ways: Fully-Dense (Exact), Diagonal Super Tile (DST)~\citep{abdulah2018exageostat},
Tile Low-Rank (TLR)~\citep{abdulah2018parallel}, and mixed-precision~\citep{abdulah2019geostatistical,abdulah2021accelerating}. The four different
implementations rely on  state-of-the-art parallel algebraic computations by exploiting advances
in algorithmic solutions and manycore computer architectures. 
The parallel implementation consists in dividing the given covariance matrix
into a set of small tiles where a single processing unit can process
a single tile at a time. The main difference between different
implementations is the structure of the underlying covariance matrix.
In dense computation, matrix tiles are represented in fully double-precision
format as shown by Figure~\ref{fig:computation_variants} (a). The provided
solution is exact but with the cost
of more computing power  and storage space. The DST implementation depends on
annihilating some of the off-diagonal tiles because their 
contributions and qualitative impact on the overall statistical problem may
be limited while depending on diagonal tiles, which should have a stronger
influence on the underlying model. Choosing the
number of tiles to be ignored is up to the user who should expect losing
some accuracy with more zero tiles.
Figure~\ref{fig:computation_variants} (b) depicts the covariance matrix in
the case of DST representation where two-diagonal tiles are represented
in dense format while all the other tiles are set to zero. The TLR implementation depends on representing
the tiles in low-rank format. Currently, \ExaGeoStat supports the Singular
Value Decomposition (SVD) technique to compress the off-diagonal tiles in
low-rank while other compression methods exist in the literature. 
The TLR computation depends on the TLR linear algebra operations that can provide
fast computation with appropriate accuracy. In Figure~\ref{fig:computation_variants} (c),
the TLR approximation method is used where the $k$ most significant 
singular values/vectors are captured for each off-diagonal tile to maintain the 
overall fidelity of the numerical model depending on the application-specific accuracy.
Finally, the mixed-precision implementation is inspired by the DST 
approximation technique. Instead of ignoring some off-diagonal tiles and setting
their elements as zero,  \ExaGeoStat represents their elements in 
lower-precision, i.e., single or half. In this
case, we can speed up the computation compared to the fully-dense implementation
but with higher accuracy than the DST approximation. Figures~\ref{fig:computation_variants} 
(d) depicts the mixed-precision covariance matrix structure supported by \ExaGeoStat.

\subsection{The ExaGeoStat Infrastructure}
To provide the different computational variants shown above, \ExaGeoStat internally
relies on three parallel linear algebra libraries to construct the basic linear algebra
operations for the MLE computation. Exact, DST, and mixed-precision approximation
computations rely on the \chameleon library, a high-performance numerical library
that provides high-performance dense linear solvers \citep{chameleonlib}
or the \dplasma library, dense linear algebra algorithms on massively parallel architectures~\citep{bosilca2011flexible}. The TLR approximation computation
depends on the \hicma library, a hierarchical linear algebra library on manycore
architectures, that provides parallel approximation solvers \citep{hicmalib}. The \hicma is
associated with the \starsh library, a high-performance $\mathcal{H}$-matrix generator
library on large-scale systems,  which provides test cases for the \hicma 
library \citep{stashlib}. Both \chameleon and \hicma libraries provide linear algebra
operations through a set of sequential task-based algorithms.

For hardware portability, \ExaGeoStat features
 the \starpu \citep{augonnet2011starpu} and \parsec \citep{bosilca2013parsec}
dynamic runtime systems as backends. The runtime system proposes a kind of abstraction
to improve the user's productivity and creativity. For example, \chameleon 
and \hicma provide sequential task-based linear algebra operations through a sequential task flow (STF) 
programming model. \starpu is able to execute the set of given sequential tasks 
in parallel with given hints of the data dependencies (e.g., read, write, and read-write). 
The main advantage of using runtime systems that rely on task-based 
implementations is becoming oblivious to the targeted hardware 
architecture. Multiple implementations of the same tasks are generated for: CPU, 
CUDA, OpenCL, OpenMP, and MPI, to name a few. To achieve the highest performance, 
the runtime system decides which implementation will achieve the highest 
performance at runtime. For the first execution, the runtime system generates 
a set of cost models that determine the best hardware for optimal performance 
during the given tasks. This set of cost models may be saved for future executions.

The top layer of ExaGeoStat is the optimization toolbox. \ExaGeoStat  relies
on an open-source {\em C}/{\em C}++ nonlinear optimization 
toolbox, \nlopt \citep{johnson2014nlopt}, to perform the MLE optimization 
operation. Among 20 global and local optimization algorithms supported 
by the \nlopt library, we selected the Bound Optimization BY Quadratic 
Approximation (BOBYQA)  algorithm to be our optimization algorithm. BOBYQA 
performs well with our target nonlinear problem with a global maximum point. BOBYQA  
is a numeric, global, derivative-free, and bound-constrained optimization 
algorithm. It generates a new computed point on each iteration by solving a 
trust-region subproblem subject to given constraints. In \ExaGeoStat, only 
upper and lower bound constraints are used. Though BOBYQA does not require 
evaluating the derivatives of the cost function, it iteratively employs an  updated
quadratic model of the objective, so there is an implicit assumption of smoothness.

\begin{figure}[h!]
  \centering
    \includegraphics[width=0.5\textwidth]{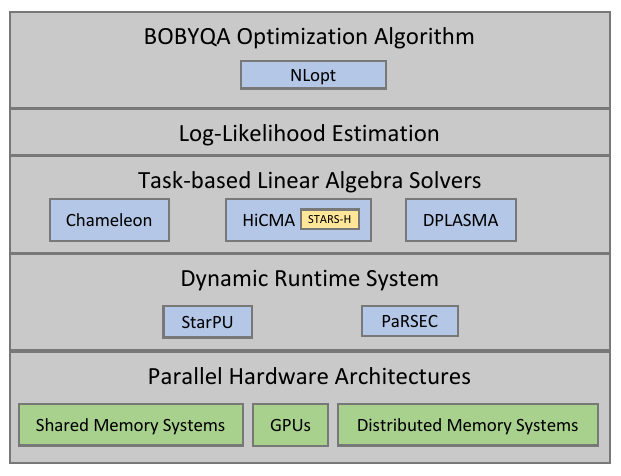}
      \caption{\ExaGeoStat infrastructure.}
  \label{fig:egs_infra}
\end{figure}

In summary, \ExaGeoStat relies on a set of software that expands the software
portability capabilities. Figure~\ref{fig:egs_infra} shows the \ExaGeoStat infrastructure
with four main layers: the BOBYQA  algorithm from the \nlopt library for optimization
purpose; the log-likelihood estimation upper-level operation;  
the \chameleon/\hicma/\dplasma libraries, which provide exact (the focus of this study) and
approximate solvers for the linear algebra operations using task-based 
parallel algorithms; and the \starpu/\parsec dynamic runtime systems, which 
translate the software for execution on the appropriate underlying hardware. Table~\ref{tab:software_dep} summarizes the complete list of software dependencies.
An optimized \blas library should be available based on the hardware 
architecture, e.g., \mkl for CPU and \openblas for GPU. The \hwloc library is 
used to abstract the underlying hardware topology to the 
runtime system. In contrast, the \gsl library provides a vast set of 
numerical functions that are needed by some of the covariance functions. 
For instance, the Bessel and gamma functions for the Mat{\'e}rn covariance function.

\subsection{{\ExaGeoStatR} Package}
To facilitate the use of large-scale executions in the \R environment, we
present a package in \R, i.e., \ExaGeoStatR \footnote{https://github.com/ecrc/exageostatR}, on
top of our \ExaGeoStat  software that provides high-performance geospatial
statistics functions in \R. This package in \R should help disseminate our software
to a large computational and spatial statistics community. To the best of
our knowledge, most of the existing \R solutions for the MLE problem are sequential
and restricted to limited data sizes. A few of them depend on running on multiple cores
within the same shared memory system by enabling \openmp support. \ExaGeoStatR targets
all the existing parallel hardware architectures, including GPUs and distributed memory
systems, with a high-level abstraction of the underlying hardware architecture
for the user. Table~\ref{tab:egsr_funs} gives an overview of current \ExaGeoStatR functions
with a description of their main operations. We provide an \exageostatr installation tutorial  in the Appendix.

\section{Simulation Studies}
\label{section3}

In this section, we provide a set of examples with associated code to
better understand the \exageostatr package. The examples possess three
goals: 1) provide step-by-step instructions of using \exageostatr on multiple
different tasks; 2) assess the performance and accuracy of the proposed exact
computation compared to existing {\em R} packages; and 3) assess
the performance of the \exageostatr package using different hardware architectures.  
All the upcoming experiments rely on the \chameleon linear algebra
library and  \starpu runtime system when executing the \exageostatr functions.

\subsection{Performance Evaluation of ExaGeoStatR}

The performance of \exageostatr is evaluated on various systems: 
the experiments in Examples 1 and 2 below are implemented on a 
Ubuntu 16.04.5 LTS workstation with a dual-socket 8-core Intel Sandy 
Bridge Intel Xeon E5-2650 without any GPU acceleration; Example 3 is 
assessed on a dual-socket 14-core Intel Broadwell Intel Xeon CPU E5-2680 v3 running 
at 2.40 GHz and equipped with 8 NVIDIA K80s (2 GPUs per board), and Example 4 is tested 
on KAUST's Cray XC40 supercomputer system, Shaheen II, with 6174 nodes, each node is dual-socket  
16-core Intel Haswell processor running at 2.30 GHz and 128 GB of DDR4 memory, and Example 5 is tested on KAUST Ibex cluster with up to 16 nodes. Two 
popular \R packages, {\geor} \citep{geoR} and {\fields} \citep{fields}, are selected as our 
references for exact computations. 

Since \exageostatr works with multiple cores and different hardware
architectures, users need to initialize their preferred settings using
the {\em exageostat\_init} function. When users want to change or 
terminate the current hardware allocation, the {\em exageostat\_finalize} function is required:

\begin{lstlisting}[language=R, texcl=true,upquote=true,literate={--}{{-{}-}}1,   basicstyle=\ttfamily,basicstyle=\footnotesize,keywordstyle=\ttfamily, deletekeywords={ beta, data,frame,length,as,character},showstringspaces=false]
> library("exageostatr")
> hardware = list (ncores = 2, ngpus = 0, ts = 320, pgrid = 1, qgrid = 1)
> exageostat_init(hardware)
> exageostat_finalize()
\end{lstlisting}

The {\em hardware = list()} specifies the required hardware to execute the code.
Here, {\em ncores} and {\em ngpus} are the numbers of CPU cores and GPU
accelerators to deploy, {\em ts} denotes the tile size used for parallelized 
matrix operations, {\em pgrid} and {\em qgrid} are the cluster topology parameters
in case of distributed memory system execution. 

\subsection{Performance Optimization Options}
In general, the performance of the \exageostatr package on shared 
memory, GPUs, and distributed memory systems can be optimized
by explicitly using \starpu optimization environment variables. For example, the
{\em STARPU\_SCHED} environment variable is used to select appropriate
parallel tasks scheduling policies provided by \starpu, such as random, eager,
and stealing. It determines how to distribute individual tasks to different processing units.
The user needs to try various schedulers to satisfy the best performance on the target hardware. Other examples of environment variables are 
{\em STARPU\_LIMIT\_MAX\_SUBMITTED\_TASKS} and {\em STARPU\_LIMIT\_MIN\_SUBMITTED\_TASKS} which control the number of submitted tasks and enable cache buffer reuse in main memory. 

\subsection{Example 1: Data Generation}
\exageostatr offers two functions to generate realizations from Gaussian random fields
with zero mean and Mat{\'e}rn covariance function shown in 
Equation \eqref{eq:matern}. The {\em simulate\_data\_exact} function 
generates a GRF at a set of irregularly spaced random locations. Five inputs 
need to be given.  The $kernel$ input accepts one of seven covariance 
functions: univariate Gaussian stationary Mat{\'e}rn - space ({\em ugsm-s}), 
bivariate Gaussian stationary flexible Mat{\'e}rn - space ({\em bgsfm-s}), 
bivariate Gaussian stationary parsimonious Mat{\'e}rn - space ({\em bgspm-s}),  
trivariate Gaussian stationary parsimonious Mat{\'e}rn - space ({\em tgspm-s}),
univariate Gaussian stationary Mat{\'e}rn - space-time ({\em ugsm-st}),
and bivariate Gaussian stationary Mat{\'e}rn - space-time ({\em bgsm-st}).
Table~\ref{tab:software_kernels}
summarizes the current covariance functions supported by \exageostatr.

The $theta$ input is a vector of the initial model parameters used to generate 
the simulated target geospatial dataset. The $theta$ vector length depends 
on the chosen $kernel$. The {\em dmetric} input accepts two values: {\em euclidean} 
for Euclidean distance or {\em great\_circle}
for great circle distance in case of spherical data \citep{veness2010calculate}. 
The $n$ input accepts the number of geospatial locations of the generated data 
in the unit square as mentioned in ~\cite{abdulah2018exageostat}. The code below 
gives a simple example of generating a realization from a univariate Gaussian stationary random field 
at 1600 random locations using the {\em simulate\_data\_exact} function:

\begin{lstlisting}[language=R, texcl=true,upquote=true,literate={--}{{-{}-}}1,   basicstyle=\ttfamily,basicstyle=\footnotesize,keywordstyle=\ttfamily, deletekeywords={ beta, data,frame,length,as,character},showstringspaces=false]
> hardware = list(ncores = 4, ngpus = 0, ts = 320, pgrid = 1, qgrid = 1)
> exageostat_init(hardware)
> data.exageo.irreg = simulate_data_exact(kernel = "ugsm-s", theta= c(1, 0.1, 0.5), 
  dmetric = "euclidean",n = 1600, seed = 0) 
> exageostat_finalize() 
\end{lstlisting}

The code starts with defining the hardware resources which will be used to run the example. The second line initiates a new \exageostatr instance. The third line simulates synthetic data using a given covariance function with a given parameter vector. Finally, the last line closes the active \exageostatr instance. The results are stored as a list, {\em data=list\{x,y,z\}}, where {\em x} and {\em y} are coordinates, and {\em z} stores the simulated realizations. Here, {\em x} and {\em y} are generated from a uniform distribution on $[0,1]$. Therefore, the generated locations are  irregular on $[0,1] \times[0,1]$ with {\em simulate\_data\_exact} function. To generate data on a regular grid, outside the range of $[0,1] \times[0,1]$, or at specific locations, one can use the {\em simulate\_obs\_exact} function by providing the coordinates {\em x} and {\em y}. The following code shows an example of generating a GRF on a $[0,2] \times [0,2]$ spatial grid with 1600 locations:


\begin{lstlisting}[language=R, texcl=true,upquote=true,literate={--}{{-{}-}}1,   basicstyle=\ttfamily,basicstyle=\footnotesize,keywordstyle=\ttfamily, deletekeywords={ beta, data,frame,length,as,character},showstringspaces=false]
> hardware = list(ncores = 2, ngpus = 0, ts = 320, pgrid = 1, qgrid = 1)
> exageostat_init(hardware)
> xy = expand.grid((1 : 40) / 20,(1 : 40) / 20)
> x = xy[,1]
> y = xy[,2]
> data.exageo.reg = simulate_obs_exact(x = x, y = y, kernel = "ugsm-s", 
  theta= c(1, 0.1, 0.5), dmetric = "euclidean")
> exageostat_finalize() 
\end{lstlisting}

\begin{figure}[h!]
\centering
\begin{subfigure}[b]{0.22\textwidth}
\centering
\includegraphics[width=\textwidth]{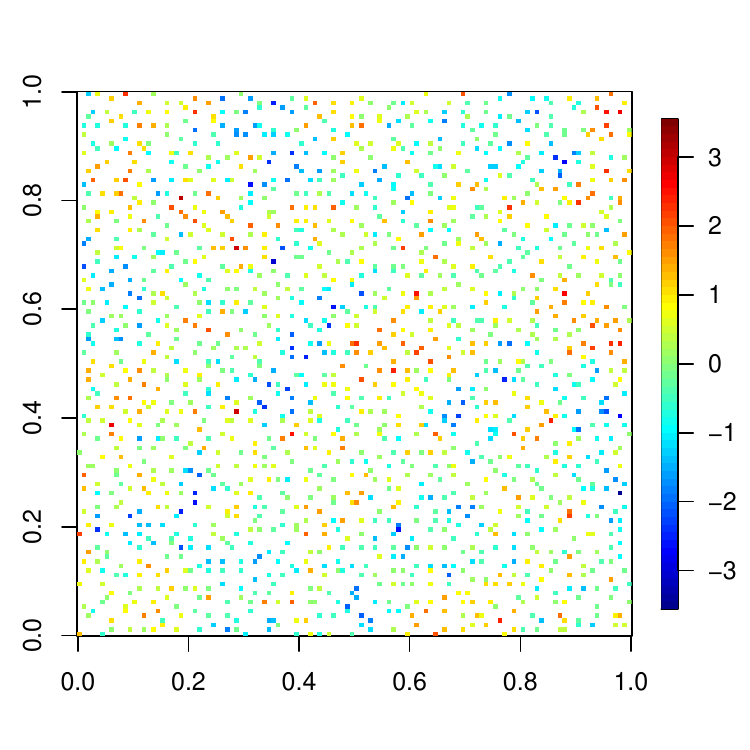}
\caption{(1, 0.03, 0.5)}
 \end{subfigure}
 \hspace{12mm}
 \begin{subfigure}[b]{0.22\textwidth}
\centering
\includegraphics[width=\textwidth]{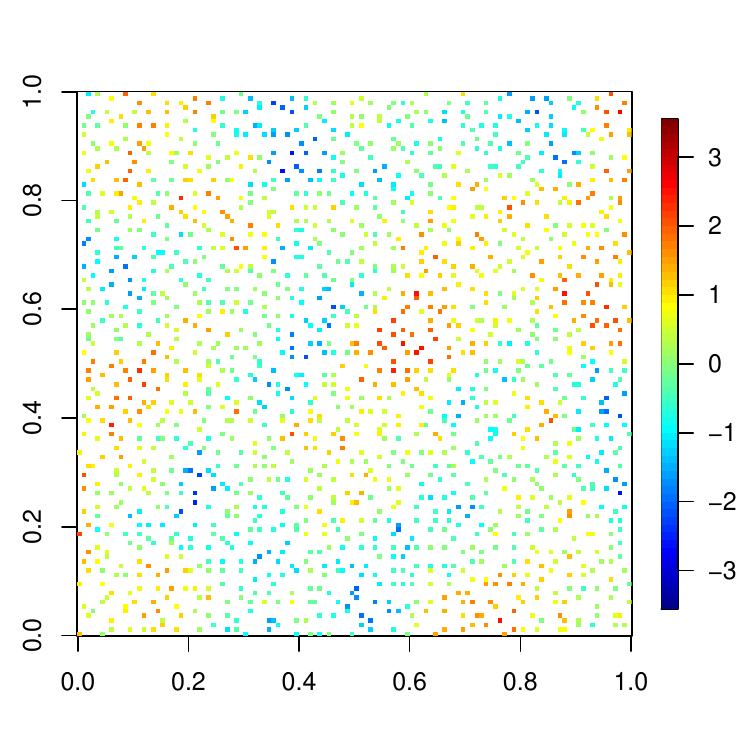}
\caption{(1, 0.1, 0.5)}
 \end{subfigure}
 \hspace{12mm}
 \begin{subfigure}[b]{0.22\textwidth}
 \centering
\includegraphics[width=\textwidth]{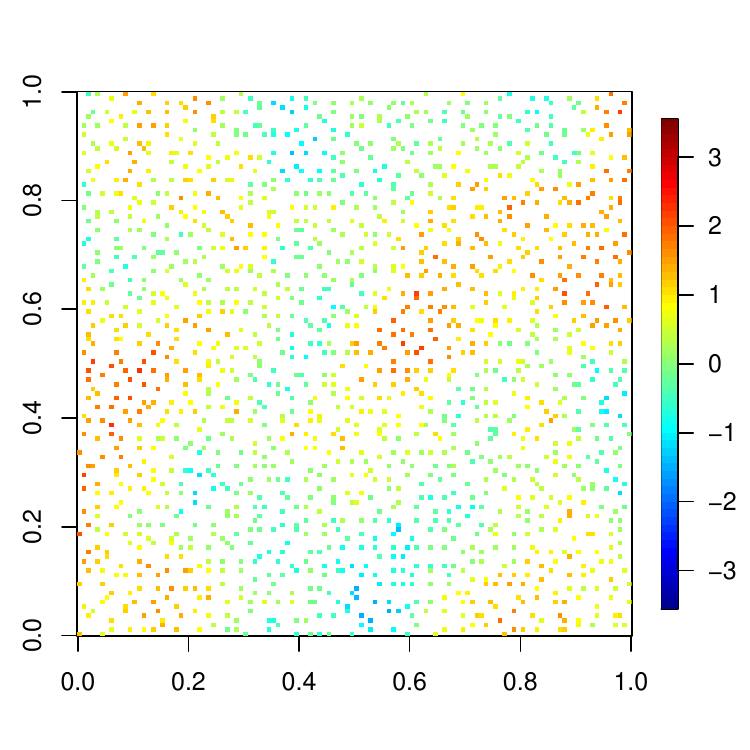}
\caption{(1, 0.3, 0.5)}
 \end{subfigure}
 
  \begin{subfigure}[b]{0.22\textwidth}
 \centering
\includegraphics[width=\textwidth]{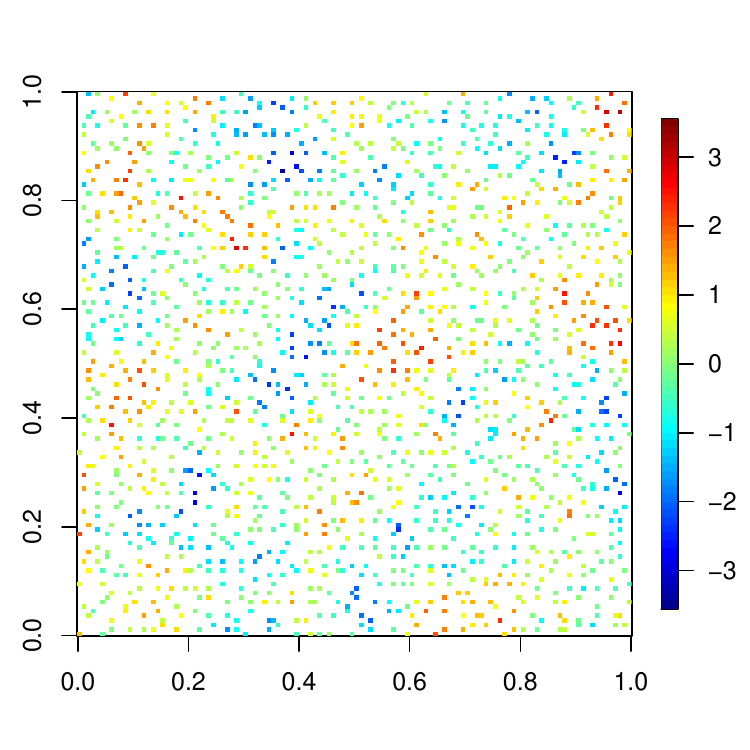}
\caption{(1, 0.03, 1)}
 \end{subfigure}
 \hspace{12mm}
  \begin{subfigure}[b]{0.22\textwidth}
  \centering
\includegraphics[width=\textwidth]{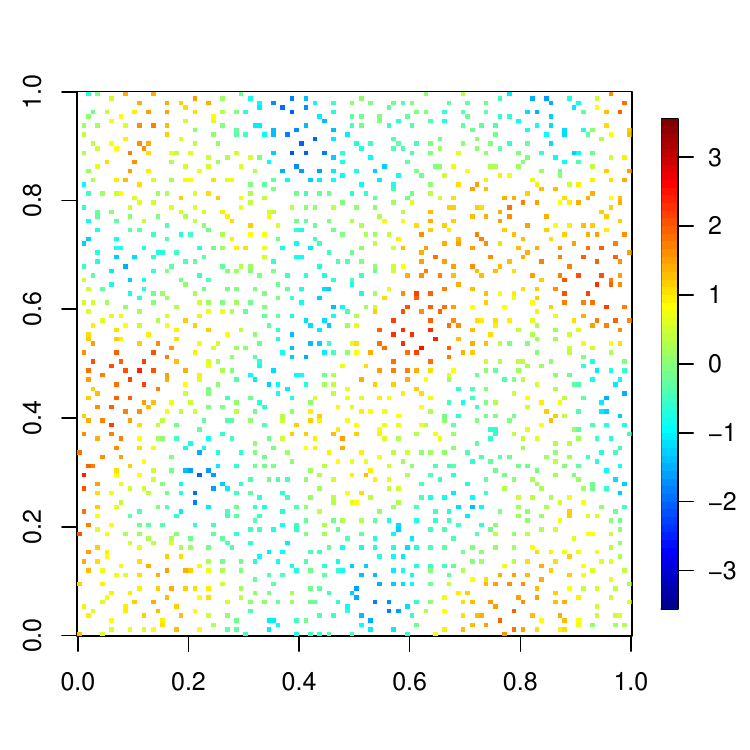}
\caption{(1, 0.1, 1)}
\label{}
 \end{subfigure}
 \hspace{12mm}
  \begin{subfigure}[b]{0.22\textwidth}
  \centering
\includegraphics[width=\textwidth]{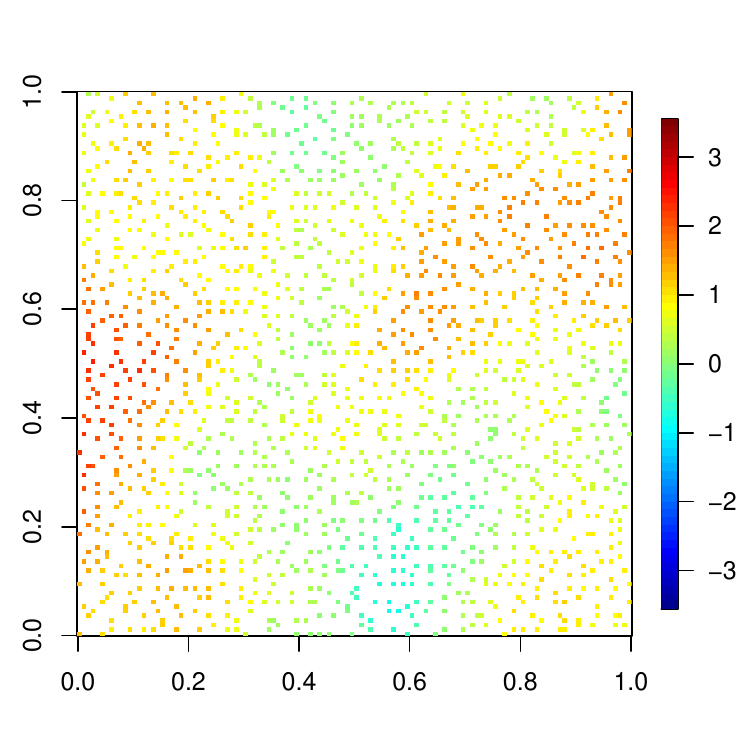}
\caption{(1, 0.3, 1)}
\label{}
 \end{subfigure}
 
 \begin{subfigure}[b]{0.22\textwidth}
  \centering
\includegraphics[width=\textwidth]{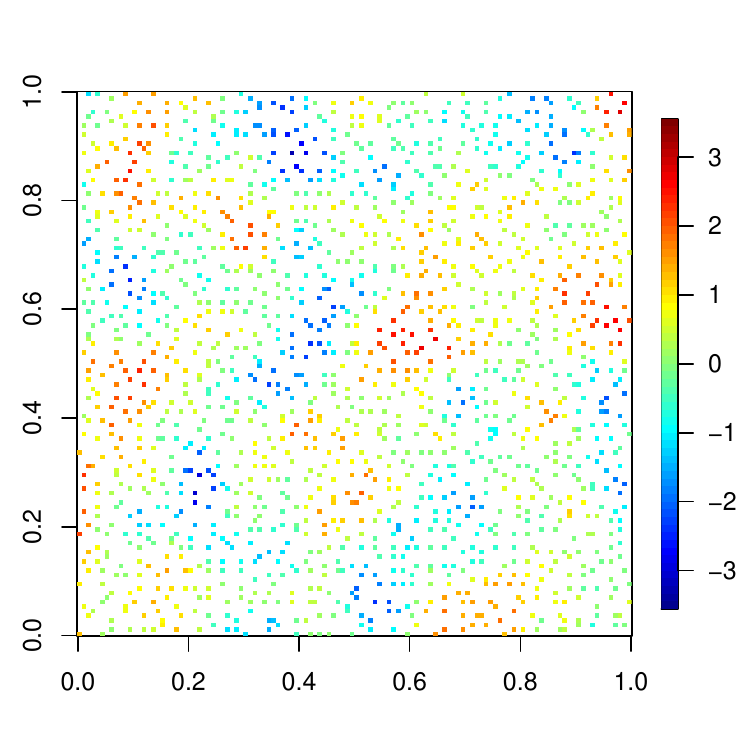}
\caption{(1, 0.03, 2)}
\label{}
 \end{subfigure}
 \hspace{12mm}
  \begin{subfigure}[b]{0.22\textwidth}
  \centering
\includegraphics[width=\textwidth]{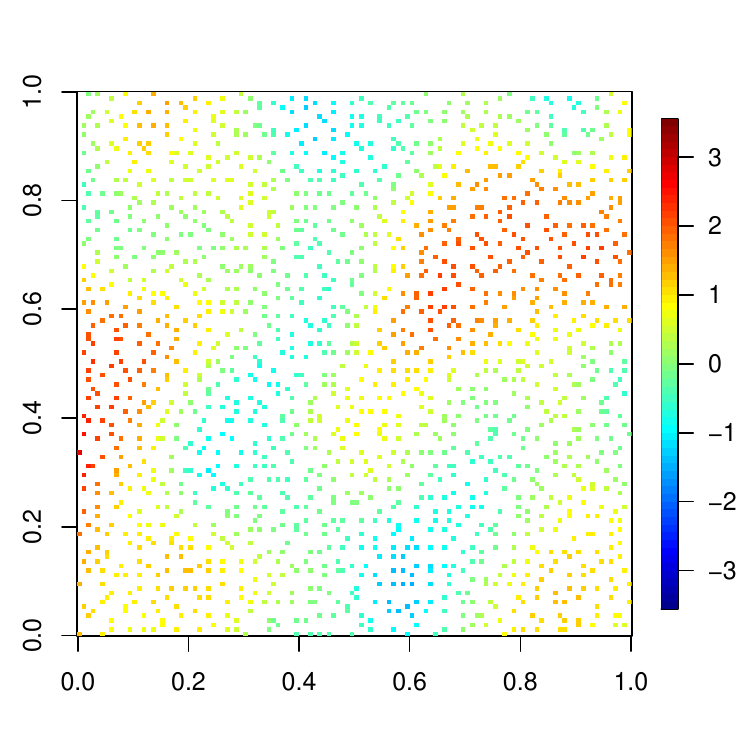}
\caption{(1, 0.1, 2)}
\label{}
 \end{subfigure}
 \hspace{12mm}
  \begin{subfigure}[b]{0.22\textwidth}
  \centering
\includegraphics[width=\textwidth]{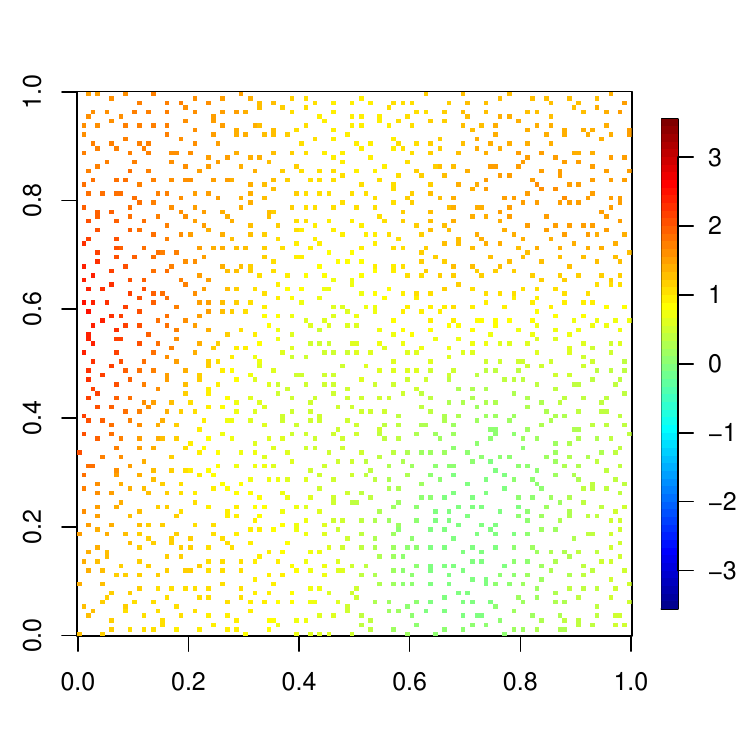}
\caption{(1, 0.3, 2)}
\label{}
 \end{subfigure}
  \caption{Data simulated using \exageostatr with univariate Mat{\'e}rn stationary covariance function in ($\sigma^2$, $\beta$, $\nu$) form at $1600$ geospatial locations in the unit square.}
   \label{fig:simulated_data}
\end{figure}

Line 5 in the code simulates synthetic data on given 2d locations using a predefined covariance function and a given parameter vector. In Figure~\ref{fig:simulated_data}, some examples of data simulated using the  \exageostatr package at $1600$ locations in the unit square using the univariate Mat{\'e}rn stationary covariance function. The data was generated using $seed = 1$.

For comparison purpose, we also show how the {\em sim.rf} function of {\fields} and the {\em grf} function of {\geor } generate similar GRFs:

\begin{lstlisting}[language=R, texcl=true,upquote=true,literate={--}{{-{}-}}1,   basicstyle=\ttfamily,basicstyle=\footnotesize,keywordstyle=\ttfamily, deletekeywords={ beta, data,frame,length,as,character},showstringspaces=false]
> library(geoR)
> sims = grf(n = 1600, grid = "reg", cov.pars = c(1, 0.1), kappa = 0.5)
> data.geoR.reg = list(x = sims$coords[, 1], y = sims$coords[, 2], z = sims$data)
\end{lstlisting}


\begin{lstlisting}[language=R, texcl=true,upquote=true,literate={--}{{-{}-}}1,   basicstyle=\ttfamily,basicstyle=\footnotesize,keywordstyle=\ttfamily, deletekeywords={ beta, data,frame,length,as,character},showstringspaces=false]
> library(fields)
> sigma_sq = 1
> grid = list(x = (1 : 40) / 20,y = (1 : 40) / 20) 
> xy = expand.grid(x=(1 : 40) / 20, y=(1 : 40) / 20)
> obj = matern.image.cov(grid = grid, theta = 0.1, smoothness = 0.5, setup = TRUE)
> sims.fields = sqrt(sigma_sq) * sim.rf(obj)
> data.fields.reg = list(x = xy[, 1], y = xy[, 2], z = c(sims.fields))
\end{lstlisting}

As can be seen, the three packages offer different types of flexibility in 
terms of data generation. However, when the goal is to generate a large GRF
on an irregular grid with more than $20$K locations, both the {\em {sim.rf}} function
and the {\em {grf}} function are not feasible. The {\em sim.rf} function simulates a
stationary GRF only on a regular grid, and the {\em {grf}} function shows memory 
issues for large size ({\em {i.e., Error:vector memory exhausted}}). On 
the other hand, the {\em {simulate\_data\_exac}t} function can easily 
generate the GRF within one minute with the following code:


\begin{lstlisting}[language=R, texcl=true,upquote=true,literate={--}{{-{}-}}1,   basicstyle=\ttfamily,basicstyle=\footnotesize,keywordstyle=\ttfamily, deletekeywords={ beta, data,frame,length,as,character},showstringspaces=false]
> n = 25600
> hardware = list(ncores = 4, ngpus = 0, ts = 320, pgrid = 1, qgrid = 1)
> exageostat_init(hardware)
> data.exageo.reg =  simulate_data_exact(kernel = "ugsm-s", theta= c(1, 0.1, 0.5), 
  dmetric = "euclidean", n, seed = 0) 
> exageostat_finalize() 
\end{lstlisting}

Simulating data on a large scale requires enough memory and 
computation resources. Thus, we recommend
the users be consistent when generating large datasets with the available 
hardware resources. We also provide a set of synthetic and real large 
spatial data examples that can be downloaded from  \url{https://ecrc.github.io/exageostat/md_docs_examples.html} for 
experimental needs.

\subsection{Example 2: Performance on Shared Memory Systems for Moderate and Large Sample Size}

\begin{figure}[t!]
 \centering
 \includegraphics[width=0.95\textwidth]{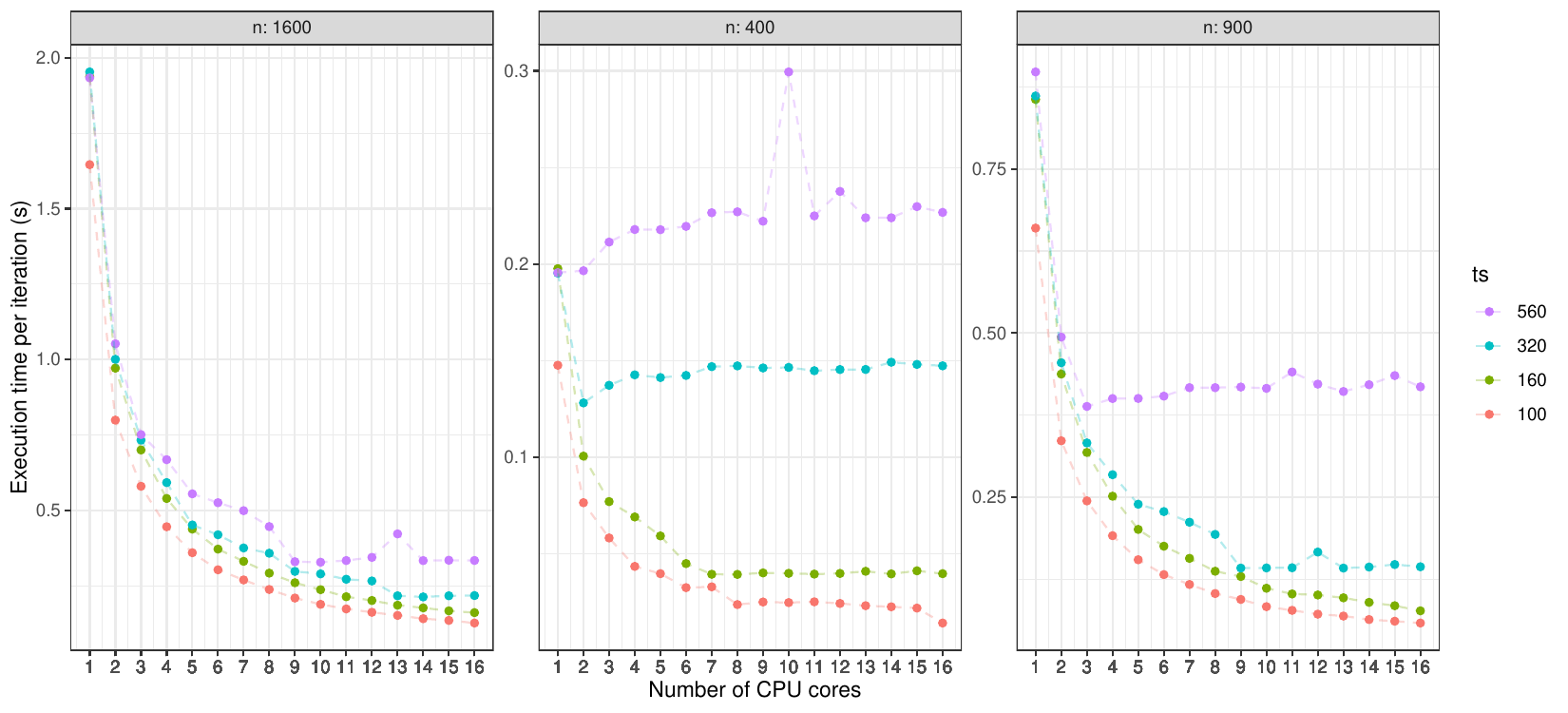}
 \caption{Parallel execution performance of \exageostatr under different hardware settings. Each subfigure corresponds to a single sample size $n$ and shows the execution time in seconds per iteration with regards to the number of cores up to 16. Curves with different colors provide the effect of tile size ({\em ts}). (Red: {\em ts=100}. Green: {\em ts=160}. Blue: {\em {ts=320}}. Purple: {\em ts=560}).}
 \label{fig:cores}
\end{figure}

To investigate the estimation of parameters based on the exact
computation, we use the {\em {exact\_mle}} function in \exageostatr. 
On a shared memory system with a moderate sample size, the number 
of cores ({\em ncores}) and tile size ({\em ts}) significantly affect the 
performance (see Figure \ref{fig:cores}). The following code shows the 
usage of the {\em {exact\_mle}} function and returns execution time per 
iteration for one combination of {\em n}, {\em ncores} and {\em ts}:

\begin{lstlisting}[language=R, texcl=true,upquote=true,literate={--}{{-{}-}}1,   basicstyle=\ttfamily,basicstyle=\footnotesize,keywordstyle=\ttfamily, deletekeywords={beta, data,frame,length,as,character,grid,kappa,list,coords,mat,time_mat,args,dim,time,time_mat,max,library}]
> hardware = list (ncores = 4, ngpus = 0, ts = 160, pgrid = 1, qgrid = 1)
> exageostat_init(hardware)
> data.exageo.reg =  simulate_data_exact(kernel = "ugsm-s", theta= c(1, 0.1, 0.5), 
  dmetric = "euclidean", n = 1600, seed = 0) 
> result = exact_mle(data.exageo.reg, kernel = "ugsm-s", dmetric = "euclidean", optimization = 
  list(clb = c(0.001, 0.001, 0.001), cub = c(5, 5, 5 ), tol = 1e-4, max_iters = 20)) 
> time = result$time_per_iter
> exageostat_finalize()
\end{lstlisting}

In the {\em {exact\_mle}} function, the first argument,
{\em {data = list\{x, y, z\}}}, is a list that defines a set of locations in
two-dimensional coordinates, {\em x and y}, and the measurement {\em z} of the 
variable of interest. The $kernel$ input defines the required 
covariance function. Here, {\em dmetric} is a distance parameter, the same 
as in the {\em {simulate\_data\_exact}} function. The {\em optimization} list 
specifies the optimization settings including the lower and upper bounds 
vectors, {\em clb} and {\em cub}, {\em tol} is the optimization tolerance 
and {\em max\_iters} is the maximum number of iterations to terminate 
the optimization process. The optimization function uses the {\em clb} vector 
as the starting point of the optimization. 

The above example has been executed using three different sample 
sizes, 16 different numbers of cores, and four different tile sizes to assess
the parallel execution performance of  \exageostatr. We visualize the results
by using the {\em ggplot} function in {\em ggplot2} \citep{Wickham:2016aa}  as shown in Figure~\ref{fig:cores}. The figure shows the computational time for the 
estimation process
using a different number of cores up to 16 cores. The y-axis shows the total
computation time per iteration in seconds, while the x-axis represents the number
of cores. The three sub-figures show the performance with different $n$ values:
400, 900, and 1600. Different curves represent different tile sizes which impact the 
performance of different hardware architectures. The figure shows that on our 
Intel Sandy Bridge machine, the best-selected tile size is 100. Tiling helps parallel execution to achieve the highest performance by increasing the reuse of data already loaded from the main memory (RAM) to the processor cache and reducing the data movements from RAM to cache. So, obtaining the best tile size depends on the processor cache size. Moving a complete tile from RAM to cache should increase the computation efficiency of the underlying algorithm. However, small tiles can also impact performance by satisfying better load balancing between the running processes, making determining the best tile size tricky. The simplest way to obtain the best tile size is to tune it on new hardware architectures before running large problems. We recommend trying different tile sizes to get the best performance from the \exageostatr package.

%
After specifying the hardware environment settings, we test the 
accuracy of the likelihood-based estimation of \exageostatr in 
comparison with {\geor } and {\fields }. The counterparts to the {\em exact\_mle} function
are {\em likfit} in {\geor } and {\em spatialProcess} in {\fields }. The simulated 
datasets from Example 1 are used as the input to assess the performance of 
parameter estimations. The synthetic datasets are generated on {\em n=1600} 
 points in $[0,1] \times [0,1]$. We take a moderate sample size 
that costs {\geor } and {\fields } approximately ten minutes to obtain enough 
results with different scenarios and iterations. The mean structure is assumed 
to be constantly zero across the region. One hundred replicates of samples are generated 
with different {\em seed} ({\em seed}$=1,\dots,100$) to quantify the 
uncertainty. We estimate the parameter values for each sample and obtain 
100 sets of estimates independently. A Mat{\'e}rn covariance kernel is selected 
to generate the covariance matrix with nine different scenarios. Specifically, the 
variance is always chosen to be one, {\em sigma\_sq = 1.0}, the spatial range 
takes three different values representing high, medium, and low spatial 
correlation, {\em beta = c(0.3, 0.1,  0.03)}, and the smoothness also takes three 
values from rough to smooth, {\em nu = c(0.5, 1, 2)}. The simulated datasets 
with moderate sample size are generated by the {\em grf} function. We 
choose the {\em grf} function in {\geor } due to its flexibility in changing the 
parameter settings and in switching between regular and irregular grids. Herein, we depend on irregular grids to generate the data.

We set the absolute tolerance to $10^{-5}$ and unset the maximum number
of iterations ({\em max\_iters = 0}) to avoid non-optimized results. Hence, each
package can show its best performance to estimate the correct value of each
parameter.  The simulated GRFs are generated by the {\em grf} function and 
stored as a list called {\em data}:

\begin{lstlisting}[language=R, texcl=true,upquote=true,literate={--}{{-{}-}}1,   basicstyle=\ttfamily,basicstyle=\footnotesize,keywordstyle=\ttfamily, deletekeywords={ beta, data,frame,length,as,character},showstringspaces=false]
> library(geoR)
> sigma_sq = 1
> beta = 0.1 # choose one from c(0.3, 0.1,  0.03)
> nu = 0.5   # choose one from c(0.5, 1, 2)
> sims = grf(n = 1600, grid = "reg", cov.pars = c(sigma_sq, beta), kappa = nu)
> data = list(x = sims$coords[,1], y = sims$coords[,2], z = sims$data)
\end{lstlisting}

We have tried to keep the irrelevant factors as consistent as possible
when comparing \exageostatr with {\geor } and {\fields }. However, we
identified some differences between the three packages that can hardly be
reconciled. For example, {\geor } estimates the mean structure together with
the covariance structure, and {\fields } does not estimate the smoothness
parameter, $\nu$, in our package. In addition, in terms of the optimization
methods, both {\geor } and {\fields } call the {\em optim} function in {\em stats}
to maximize the likelihood function. The {\em optim} function includes six methods
such as {\em Nelder-Mead} and {\em BFGS}. However, \exageostatr uses the 
BOBYQA algorithm, which is one of the optimization algorithms of the 
sequential \nlopt library in {\em C/C++}. The {\em BOBYQA} algorithm has 
the best performance in terms of MLE estimation. However, it is not 
available in the {\em {optim}} function. Table \ref{tab:packages} lists the 
differences between the three packages.

Multiple algorithms are offered by the {\em optim} function
and further implemented by {\geor } and {\fields }. However, many 
studies in the literature point out that the {\em optim} function is 
not numerically stable for a large number of mathematical functions, especially 
when a re-parameterization exists \citep{mullen2014continuous, nash2011unifying, nash2014best}. Based on the 100 simulated samples, we show that \exageostatr provides 
faster computation and gives more accurate and robust estimations 
with regards to the initial value and grid type.


We first estimate the parameters by {\em exact\_mle} in \exageostatr. We
use the number of cores to be eight to reproduce the results on most machines.
Users can specify their settings and optimize the performance by referring to 
the results in Figure \ref{fig:cores}. The final results also report the time per iteration, total 
time, and the number of iterations for each optimization:


\begin{lstlisting}[language=R, texcl=true,upquote=true,literate={--}{{-{}-}}1,   basicstyle=\ttfamily,basicstyle=\footnotesize,keywordstyle=\ttfamily, deletekeywords={ beta, data,frame,length,as,character},showstringspaces=false]
> hardware =  list (ncores = 8, ngpus = 0, ts = 100, pgrid = 1, qgrid = 1)
> exageostat_init(hardware)
> result = exact_mle(data, kernel = "ugsm-s", dmetric = "euclidean", optimization =
      list(clb = c(0.001, 0.001, 0.001), cub = c(5, 5, 5 ), tol = 1e-4, max_iters = 20))
> para_mat  = result$est_theta
> time_mat = c(result$time_per_iter, result$total_time, result$no_iters)
> exageostat_finalize()
\end{lstlisting}

Then we estimate the parameters under the same scenarios using
the {\em likfit} function in {\geor } and the {\em spatialProcess} function
in {\fields }. For {\geor } and {\fields }, the chosen optimization 
options are {\em method = c("Nelder-Mead")}, {\em abstol = 1e-5}, and 
{\em maxit = 500}, where the maximum number of iterations set as 
500 could never be reached. Because {\fields } does not optimize the smoothness, $\nu$, so we set it as the true value (an advantageous favor for {\fields }). {\geor } has to optimize the 
mean parameter, at least a constant, but it is treated to be independent of the 
covariance parameters as the mean value of data. In addition, to accelerate the 
optimization of {\fields }, we minimize the irrelevant computation by setting {\em gridN = 1}:

\begin{lstlisting}[language=R, texcl=true,upquote=true,literate={--}{{-{}-}}1,   basicstyle=\ttfamily,basicstyle=\footnotesize,keywordstyle=\ttfamily, deletekeywords={ beta, data,frame,length,as,character},showstringspaces=false]
> result = spatialProcess(x = cbind(data$x,data$y), y = data$z, 
    cov.args = list(Covariance = "Matern", smoothness = nu), gridN = 1,  reltol = 1e-05)    
> para_mat = c( result$MLESummary[[7]],result$MLESummary[[8]], result$args$smoothness)
> time_mat = c(result$timingTable[3,2]/dim(result$MLEJoint$lnLike.eval)[1],
              result$timingTable[3,2], dim( result$MLEJoint$lnLike.eval)[1])

\end{lstlisting}


Using the \geor package:

\begin{lstlisting}[language=R, texcl=true,upquote=true,literate={--}{{-{}-}}1,   basicstyle=\ttfamily,basicstyle=\footnotesize,keywordstyle=\ttfamily, deletekeywords={ beta, data,frame,length,as,character},showstringspaces=false]
> time = system.time( fit_obj <- likfit( coords = cbind(data$x, data$y), 
    data = data$z, trend = "cte", ini.cov.pars = c(0.001, 0.001), fix.nugget  = TRUE, 
    nugget = 0, fix.kappa = FALSE, kappa = 0.001, cov.model = "matern", lik.method = "ML",
    limits =  pars.limits(sigmasq = c(0.001, 5), phi = c(0.001, 5), kappa = c(0.001, 5)), 
    method = "Nelder-Mead", control = list(abstol = 1e-5, maxit = 500)))[3]
> time_mat = c(time / fit_obj$info.minimisation.function$counts[1],
                   time, fit_obj$info.minimisation.function$counts[1])
> para_mat = c(fit_obj$sigmasq, fit_obj$phi, fit_obj$kappa)
\end{lstlisting}

The computational efficiency is compared based on the average execution
time per iteration and the average number of iterations. As shown 
in Table \ref{tab:iters}, the running time per iterations of \exageostatr is about 
12X and 7X faster than {\geor } and {\fields }, respectively. The running 
time per iteration is robust between different scenarios. Since {\fields } does 
not estimate the smoothness parameter, it runs faster
than {\geor }. Although {\geor } also estimates an extra constant mean 
parameter, it does not affect the computation much because the mean 
parameter is simply the mean of the measurements {\em z} and is estimated 
separately. Table \ref{tab:iters} also shows the number of iterations to reach the tolerance. 
We can see that \exageostatr requires more iterations but much less time to reach the accuracy.


\begin{figure}[]
 \centering
 \includegraphics[width=0.8\textwidth]{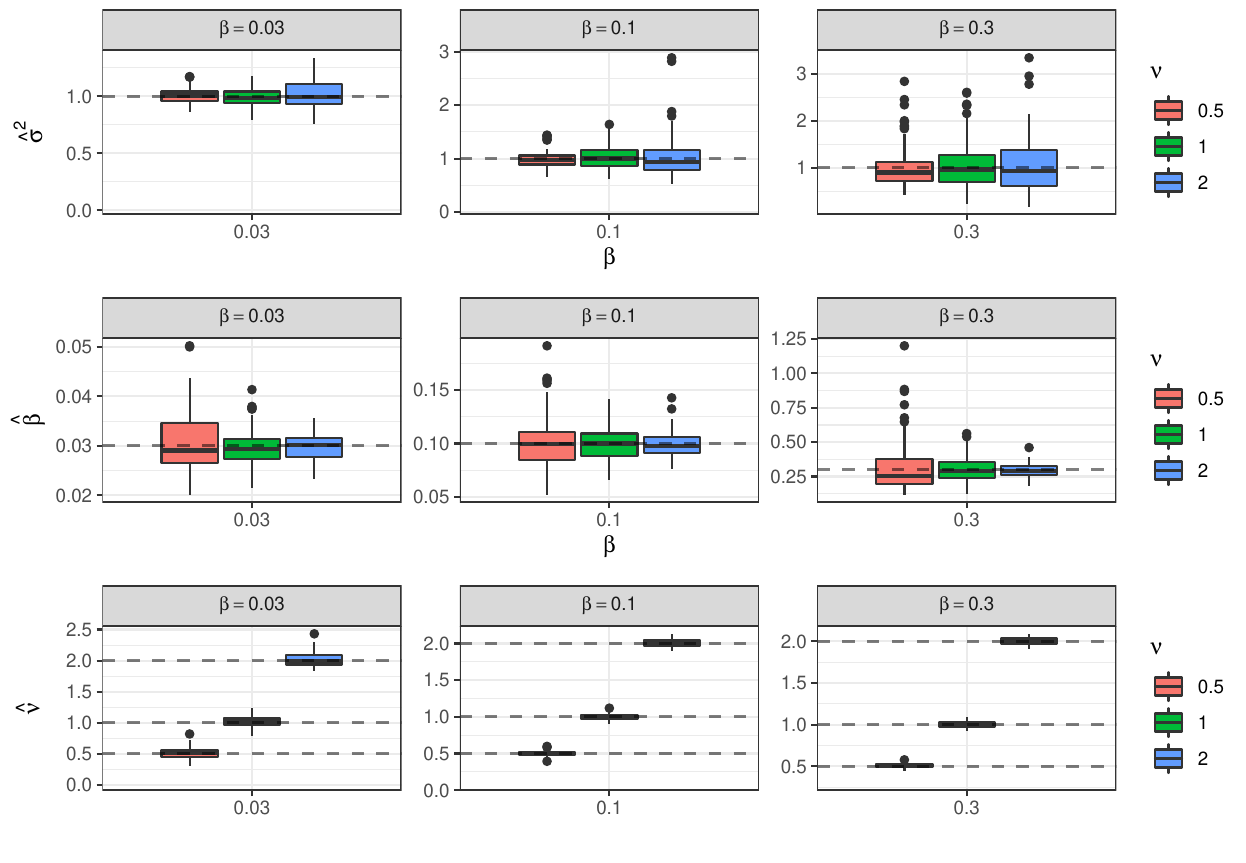}
   \caption{The estimation accuracy of \exageostatr with different sets of parameter vectors. Each row shows the estimation of a parameter among variance, $\sigma^2$, spatial range, $\beta$, and smoothness, $\nu$. Each column corresponds to one setting of spatial range, $\beta$, and the color of the boxplots identifies a type of smoothness, $\nu$. True values are represented by a horizontal dashed line.}
  \label{fig:estimation1}
\end{figure}
\begin{figure}[]
 \centering
  ~\includegraphics[width=0.8\textwidth]{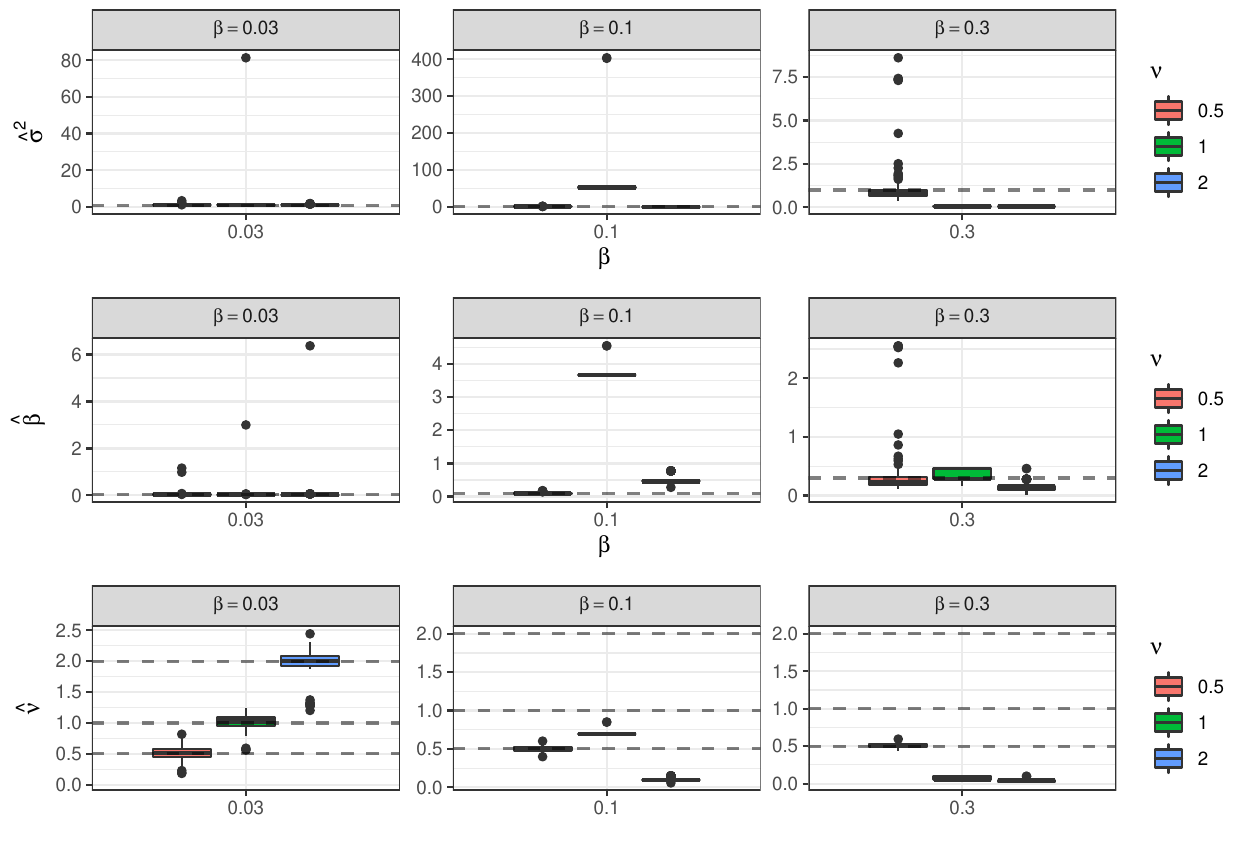}
    \caption{The estimation accuracy of {\geor } with different sets of parameter vectors.}
  \label{fig:estimation2}
\end{figure}
  \begin{figure}[h!]
 \centering
   \includegraphics[width=0.8\textwidth]{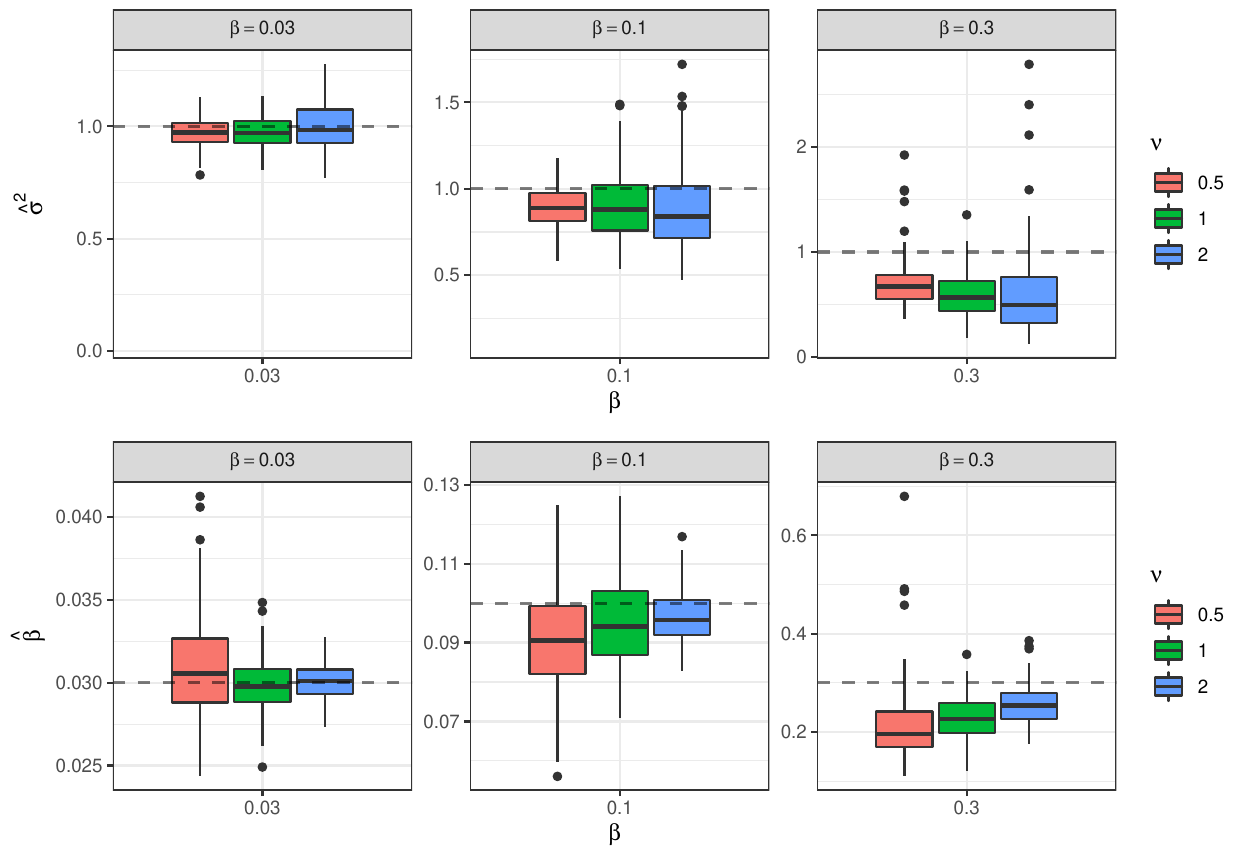}
  \caption{The estimation accuracy of {\fields} with different sets of parameter vectors. The results related to {\fields } only have two rows since the package does not estimate $\nu$.}
 \label{fig:estimation3}
\end{figure}

Figures~\ref{fig:estimation1},~\ref{fig:estimation2}, and~\ref{fig:estimation3} show the estimation accuracy between the
three packages using boxplots. It is clear that \exageostatr outperforms
{\geor } and {\fields }. Together with Table \ref{tab:iters}, we can see 
that \exageostatr requires more iterations when $\nu$ increases since we 
set the initial values to be 0.001 for all scenarios. It implies that even under 
the circumstance of bad initial values, e.g., $\nu=2$ and $\beta=0.3$, \exageostatr can
 reach the global maximum by taking more iterations. However, the estimation 
 performance of both {\geor } and {\fields } becomes worse when the initial values 
 deviate from the truth. In particular, {\geor } reaches a local maximum only after 
 about 20 iterations for medium and large smoothness and spatial range, so that its 
 estimated values are even not inside the range of {\fields } and \exageostatr. 

The result is mainly due to the numerical optimization of $\nu$, which 
involves the non-explicit Bessel function in the Mat{\'e}rn kernel. Therefore, {\fields } has 
a more robust estimation because it does not estimate the 
smoothness (we fix it as the truth), although {\fields } calls the same {\em optim} function 
as {\geor}. However, even though \exageostatr estimates the smoothness 
parameter, our package still outperforms {\fields } in terms of $\beta$ and $\sigma^2$, 
especially for medium and large spatial range correlation.

Other optimization methods rather than ``Neader-Mead'' are also explored, 
such as the default optimization option of  {\fields }, ``BFGS''. As a quasi-Newton 
method, ``BFGS'' is fast but not stable in many cases. Similar to the worst result
of {\geor }, the optimization jumps out after only a few steps and reports incorrect 
results even with a decent guess of initial values. Even worse, 
both {\geor } and {\fields }  report errors in computing the inverse of the 
covariance matrix ({\em i.e., Error:error in solve.default(v\$varcov, xmat):system 
is computationally singular}).

The experiment above shows that the {\em optim} function used by 
both {\geor } and {\fields } is not stable with regards to the initial values, no matter
what algorithm we choose. In addition, by simulating GRFs on an irregular 
grid ({\em grid = "irreg"}), few simulated GRFs can get the output without any error. For data
on an irregular grid, locations can be dense so that the distances between certain points
are very close to each other. Therefore, the columns of the covariance matrix 
corresponding to the dense locations can be numerically equal. For the data on a unit 
square, we find that \exageostatr may only have the singularity problem when the 
closest distance is less than {\em 1e-8}. On the contrary, the problem occurs when the 
smallest distance is close to {\em 1e-4} for {\fields } and {\geor }. As a 
result, although \exageostatr is designed to provide a faster computation by making 
use of manycore systems, the 
optimization algorithm it is based upon gives a more accurate and robust estimation than 
any algorithm in the {\em optim} function.

\begin{figure}[h!]
 \centering
 \includegraphics[width=0.9\textwidth]{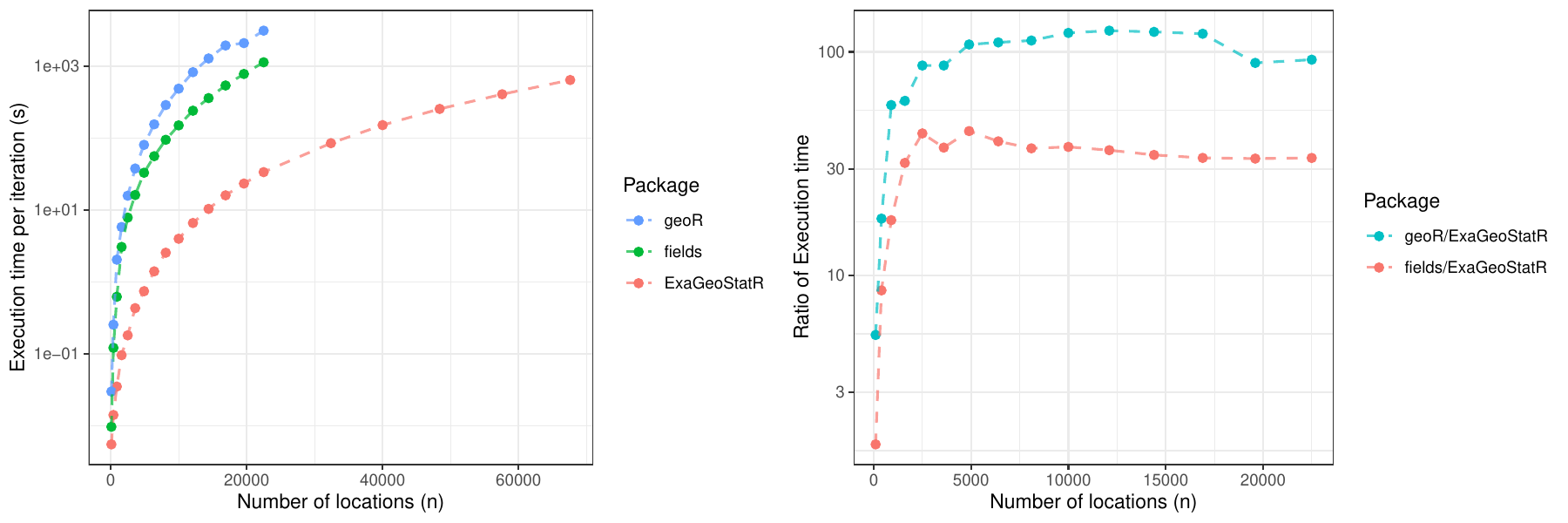}
 \caption{The execution time per iteration as $n$ increases for {\geor },  {\fields }, and  \exageostatr with 8 CPU cores. The covariance parameters are set to be $(\sigma^2,\beta,\nu)=(1,0.1,0.5)$. Each curve on the left panel shows the exact running time per iteration of one package, and each curve on the right panel gives the ratio of execution time compared to \exageostatr. The $y$-axis is shown in $\log_{10}$ scale.}
 \label{fig:runtime}
\end{figure}

We also investigate the computational time of the three packages as $n$ increases. The hardware settings remain the same for \exageostatr with 8 CPU cores. The max number of iterations 
is set to 20 for all three packages to accelerate the estimation. The results report the total 
computational time. The tested number of locations ranges from 100 to 90,000. However, both {\geor } and {\fields } take too much time when $n$ is large. For example, the estimation with {\geor } 
for 22,500 locations requires more than 17 hours. Thus we stop the simulation for {\geor } and {\fields } at the size of 22,500 and only show the execution time of \exageostatr for larger $n$ with 8 
CPU cores. The results are shown in Figure \ref{fig:runtime}. It can be seen that, when $n = 22,500$, \exageostatr is 33 times faster than {\fields } and 92 times faster than {\geor }. 

\subsection{Example 3: Extreme Computing on GPU Systems}
Since the main advantage of \exageostatr is the multi-architecture compatibility, 
here we provide another example of using \exageostatr 
on GPU systems. We choose the number of locations ranging from 1600 
to approximately 100K. The {\em R} code below shows an example of how 
to use 26 CPU cores and 1 GPU core with $n=25{,}600$:


\begin{lstlisting}[language=R, texcl=true,upquote=true,literate={--}{{-{}-}}1,   basicstyle=\ttfamily,basicstyle=\footnotesize,keywordstyle=\ttfamily, deletekeywords={ beta, data,frame,length,as,character},showstringspaces=false]
> n = 25600 
> hardware =  list (ncores = 26, ngpus = 1,  ts=960, pgrid=1, qgrid=1)
> exageostat_init(hardware)
> data = simulate_data_exact(kernel = "ugsm-s", theta = c(1, 0.1, 0.5),
  dmetric = "euclidean", n, seed = 0)
> result_cpu = exact_mle(data, kernel = "ugsm-s", dmetric = "euclidean", optimization = 
  list(clb = c(0.001, 0.001, 0.001), cub = c(5, 5, 5 ), tol = 1e-4, max_iters = 20))
> time_cpu = result_cpu$time_per_iter
> exageostat_finalize()
\end{lstlisting}

\begin{figure}[h!]
 \centering
 \includegraphics[width=0.7\textwidth]{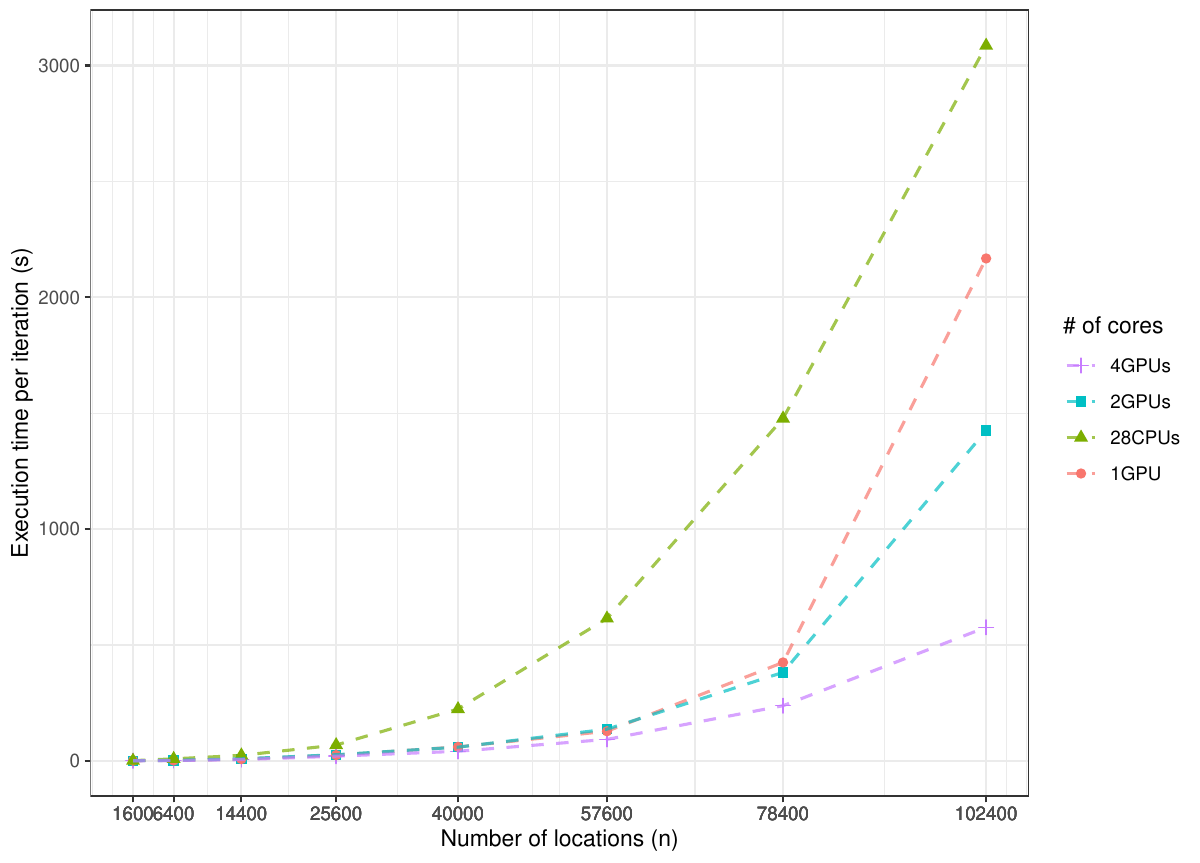}
 \caption{Execution performance of \exageostatr under different CPU and GPU combinations. The covariance parameters are set to be $(\sigma^2,\beta,\nu)=(1,0.1,0.5)$. Each curve corresponds to the execution time per iteration with regards to different sample sizes, $n$.}
 \label{fig:gpu}
\end{figure}

The above code shows that \exageostatr has a user-friendly interface to 
abstract the underlying hardware architecture to the user. The user needs 
only to specify the number of cores and GPUs required for one's execution. 
Figure \ref{fig:gpu} reports the performance with different numbers of GPU 
accelerators: 1, 2, and 4. The figure also shows the curve using the maximum 
number of cores (28-core) on the machine without GPU support. The figure 
demonstrates how GPUs speed up the execution time compared to CPU-based 
running. Moreover, the figure shows the scalability using different numbers of GPUs.


\subsection{Example 4: Extreme Computing on Distributed Memory Systems}

This subsection gives an example of using \exageostatr on distributed
memory systems (i.e., supercomputer Shaheen II Cray XC40). As for shared memory and 
GPU systems, the \exageostatr package abstracts the underlying hardware to a 
set of parameters. With distributed systems, the user needs to define four 
main parameters: {\em pgrid} and {\em qgrid} which represent a set of  
nodes arranged in a {\em pgrid} $\times$ {\em qgrid} rectangular array of 
nodes (i.e, two-dimensional block-cyclic distribution), {\em ncores} which 
represents the number of physical cores in each node, and {\em ts} which 
represents the optimized tile size. Another example of the usage of \exageostatr 
on a distributed memory system  with 31 CPU cores, $4 \times 4$ rectangular 
array of nodes, {\em ts=960}  and,  {\em n=250,000} is shown below:


\begin{lstlisting}[language=R, texcl=true,upquote=true,literate={--}{{-{}-}}1,   basicstyle=\ttfamily,basicstyle=\footnotesize,keywordstyle=\ttfamily, deletekeywords={ beta, data,frame,length,as,character},showstringspaces=false]
> n = 250000 
> hardware = list (ncores = 31, ts=960,  pgrid=4, qgrid=4)
> exageostat_init(hardware)
> data = simulate_data_exact(kernel = "ugsm-s", theta = c(1, 0.1, 0.5),
  dmetric = "euclidean", n, seed = 0)
> result_cpu = exact_mle(data, result_cpu = exact_mle(data, kernel = "ugsm-s", 
  dmetric = "euclidean", optimization = list(clb = c(0.001, 0.001, 0.001),
  cub = c(5, 5, 5 ), tol = 1e-4, max_iters = 20)
> time_cpu = result_cpu$time_per_iter
> exageostat_finalize()
\end{lstlisting}

\begin{figure}[b!]
 \centering
 \includegraphics[width=0.7\textwidth]{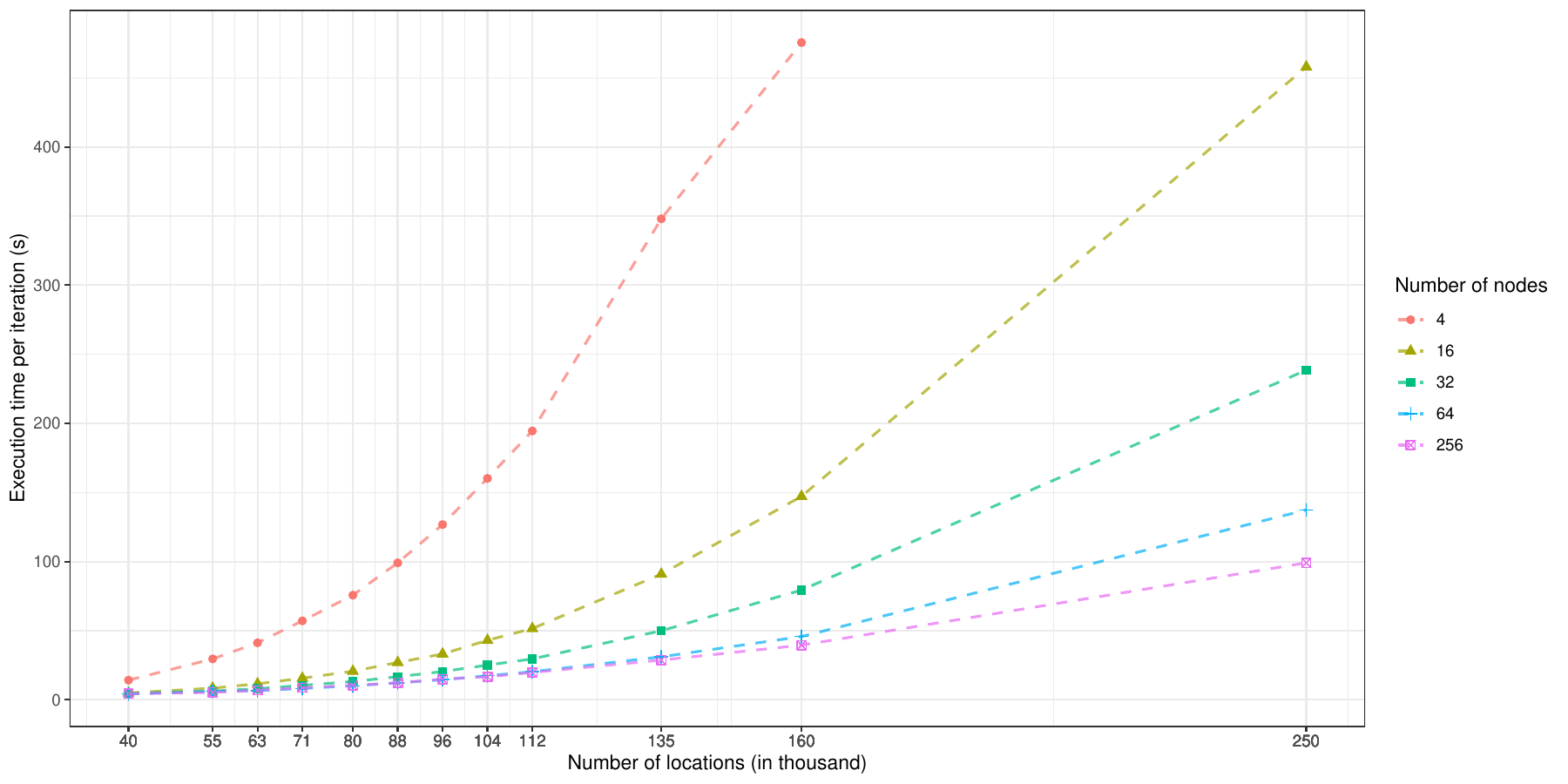}
 \caption{Performance of \exageostatr using different numbers of nodes. The time per iteration is averaged over $20$ iterations. The realization is generated from a zero-mean GRF under the Mat{\'e}rn covariance structure with the parameters $(\sigma^2,\beta,\nu)=(1,0.1,0.5)$.}
 \label{fig:shaheen}
\end{figure}

Figure~\ref{fig:shaheen} shows the performance results of running
different problem sizes on Shaheen II Cray XC40 using different
numbers of nodes. The distribution of the nodes 
are $2\times2$, $4\times4$, $8\times4$, $8\times8$, and $16\times16$. The figure 
shows strong scalability of \exageostatr with different numbers of nodes up to 64 nodes. 
The reported performance is the time per iteration averaged over $20$ iterations with settings:
\begin{lstlisting}[language=R, texcl=true,upquote=true,literate={--}{{-{}-}}1,   basicstyle=\ttfamily,basicstyle=\footnotesize,keywordstyle=\ttfamily, deletekeywords={ beta, data,frame,length,as,character},showstringspaces=false]
> export STARPU_SCHED = eager.
> export STARPU_LIMIT_MIN_SUBMITTED_TASKS = 9000.
> export STARPU_LIMIT_MAX_SUBMITTED_TASKS = 10000.

 \end{lstlisting}


\subsection{Example 5: \exageostatr Versus \biggp on Distributed Systems}
This subsection gives an example of using \exageostatr on the Ibex HPC cluster from KAUST (\href{https://www.hpc.kaust.edu.sa/ibex}{https://www.hpc.kaust.edu.sa/ibex}) using up to 16 
40-core Intel Skylake and 40-core Intel Cascadelake nodes. We intend in this example to compare \exageostatr with \biggp from a performance perspective on a distributed system. We focus on the performance of the
Cholesky factorization operation since it is the most time-consuming operation when evaluating the likelihood function. We rely on the Mat{\'e}rn covariance function to build the target covariance matrix.

As in Example 4,  the user needs to define four  main parameters to be able
to run \exageostatr script on distributed systems: 
{\em pgrid} and {\em qgrid} which represent a set of  
nodes arranged in a {\em pgrid} $\times$ {\em qgrid} rectangular array,
{\em ncores} which represents the number of cores used in each node,
and {\em ts}, which represents the tuned tile size. 

The \biggp package uses $RMPI$~\citep{RMPI}, an MPI interface in R, to perform the GP modeling
in a distributed manner. Below is an example of \biggp code to calculate the Cholesky
factorization for a given matrix that can be executed on distributed-memory systems:

\begin{lstlisting}[language=R, texcl=true,upquote=true,literate={--}{{-{}-}}1,   basicstyle=\ttfamily,basicstyle=\footnotesize,keywordstyle=\ttfamily, deletekeywords={ beta, data,frame,length,as,character},showstringspaces=false]
> library("bigGP")
> p = 4
> bigGP.init(p-1)                 
> m = 100                           
> gd <- seq(0, 1, length = m)
> locs = expand.grid(x = gd, y = gd)
> theta <- c(1, 0.1, 0.5)
> mpi.bcast.Robj2slave(theta)
> mpi.bcast.Robj2slave(covfunc)
> mpi.bcast.Robj2slave(locs)
> mpi.bcast.cmd(indices <- localGetTriangularMatrixIndices(nrow(locs)))
> mpi.bcast.cmd(C <- covfunc(theta, locs, indices))
> remoteLs()                
> remoteCalcChol('C', 'L', n = m^2)
> bigGP.quit()

\end{lstlisting}
Here $p$ represents the total number of nodes to run the script. \biggp relies on one master process and $p -1$ slave processes. The user has to submit the jobs with $p$ processes. In line 3, a \biggp instance is initiated using 3 slaves nodes. $n$ represents the number geospatial locations where $n = m^2$. $locs$ is the set of locations, and $theta$ is the parameter vector to use with a predefined covariance function kernel. The  $remoteCalcChol$ function is the Cholesky factorization function in the \biggp package.

\begin{figure}[]
 \centering
  \begin{subfigure}{0.48\textwidth}
   \centering
 \includegraphics[width=1\textwidth]{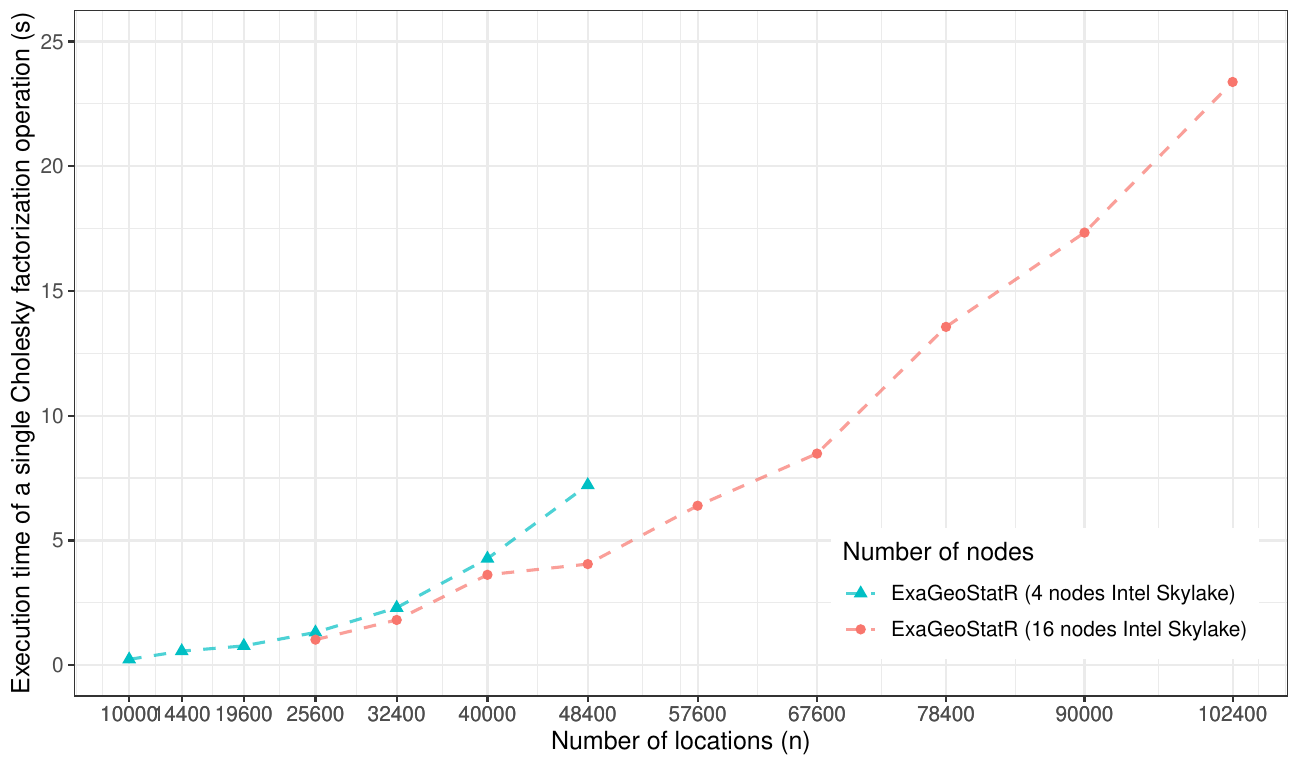}
 \subcaption{\ExaGeoStatR on Intel Skylake}
 \end{subfigure}
  \begin{subfigure}{0.48\textwidth}
   \centering
  \includegraphics[width=1\textwidth]{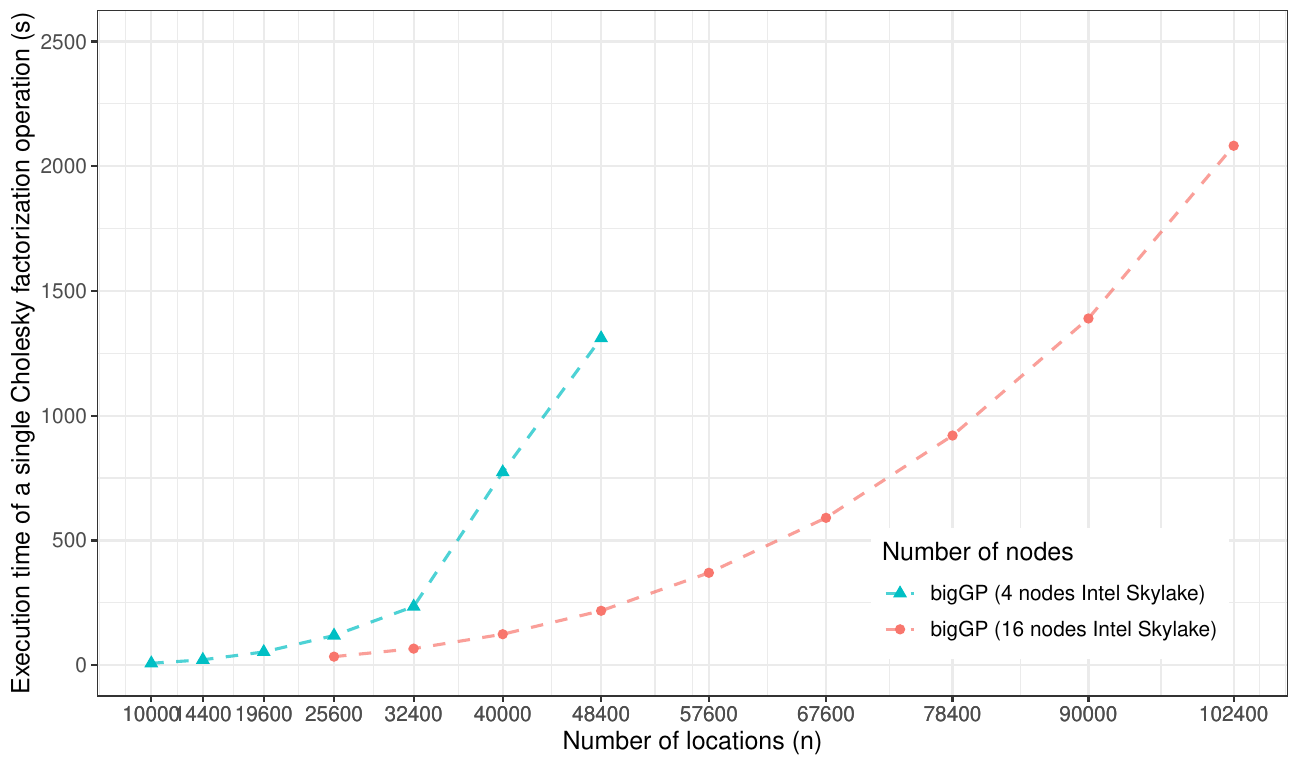}
   \subcaption{\biggp on Intel Skylake}
   \end{subfigure}
  \begin{subfigure}{0.48\textwidth}
   \centering
 \includegraphics[width=1\textwidth]{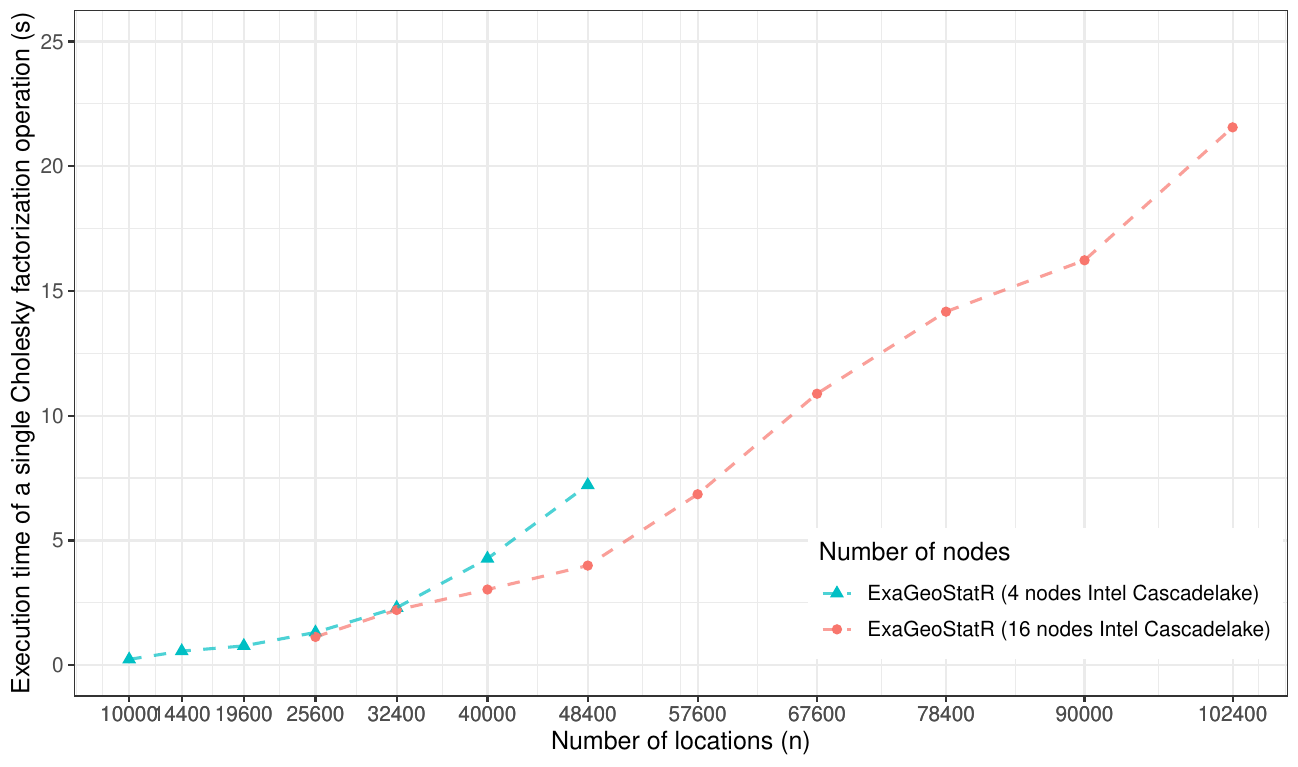}
 \subcaption{\ExaGeoStatR on Intel Cascadelake}
 \end{subfigure}
  \begin{subfigure}{0.48\textwidth}
   \centering
  \includegraphics[width=1\textwidth]{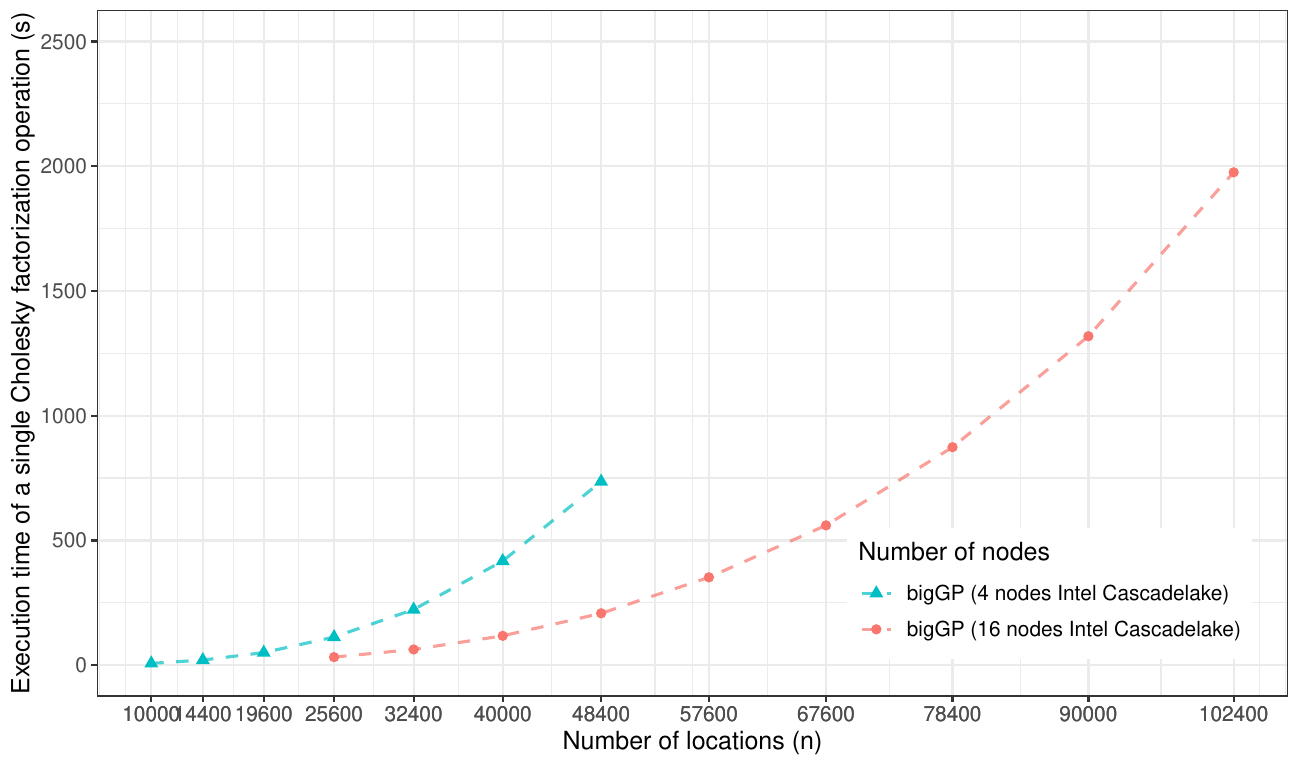}
   \subcaption{\biggp on Intel Cascadelake}
   \end{subfigure}
 \caption{Performance of the Cholesky factorization operation using  \exageostatr (a-c) and  \biggp (b-d) packages on 4 and 16 nodes on the KAUST Ibex HPC Cluster, where each node is a 40-core Intel Skylake/Cascadelake processors.}
  \label{fig:biggpvsexageostatr}
\end{figure}



Due to the high network congestion of the cluster, we repeated each run five times and selected the best execution time for both \exageostatr and \biggp. Figure \ref{fig:biggpvsexageostatr}  shows the execution time for running a single Cholesky factorization on 4 and 16 Intel Skylake/Cascadelake nodes using the Ibex cluster. The figure shows how both \exageostatr and \biggp can take advantage of increasing the number of nodes to improve the overall performance of the Cholesky factorization. On both architectures, it is demonstrated that \exageostatr outperforms \biggp using a different number of nodes which shows the benefits of tile-based parallel solvers and runtime systems compared to the block-based parallel solvers and pure OpenMP/MPI implementation. 

\section{Application to Sea Surface Temperature Data: A tutorial}
\label{section4}
West-blowing trade winds in the Indian Ocean push warm surface waters against
the eastern coast of Africa. These waters move south along the coastline, eventually
spilling out along the Indian and Atlantic Oceans boundary. This jet of warm water, known
as the Agulhas Current, collides with the cold, west-to-east-flowing Antarctic Circumpolar 
Current, producing a dynamic series of meanders and eddies as the two waters mix. The 
result makes for an interesting target for spatial analysis that we illustrate as a tutorial.

This application study provides an example where the MLE is computed in
high dimensions, and \ExaGeoStatR facilitates the procedure on many-core systems. 
We use the sea surface temperature collected by satellite for the Agulhas and surrounding
areas off the shore of South Africa. The data covers 331 days, from 
January 1 to November 26, 2004. The region is abstracted into a $72 \times 240$ regular 
grid, with the grid lines denoting the latitudes and longitudes and the spatial resolution is 
approximately $25$ kilometers, though exact values depend on 
latitude. Although {\em fields} and {\em geoR} do not have input dimension limits, the 
computation with \ExaGeoStatR has a distinct advantage on parallel architectures, hence 
more suitable for MLE with more optimization iterations to reach convergence.

\begin{figure}[t!]
 \centering
 \begin{subfigure}{0.8\textwidth}
 \includegraphics[width=0.99\linewidth]{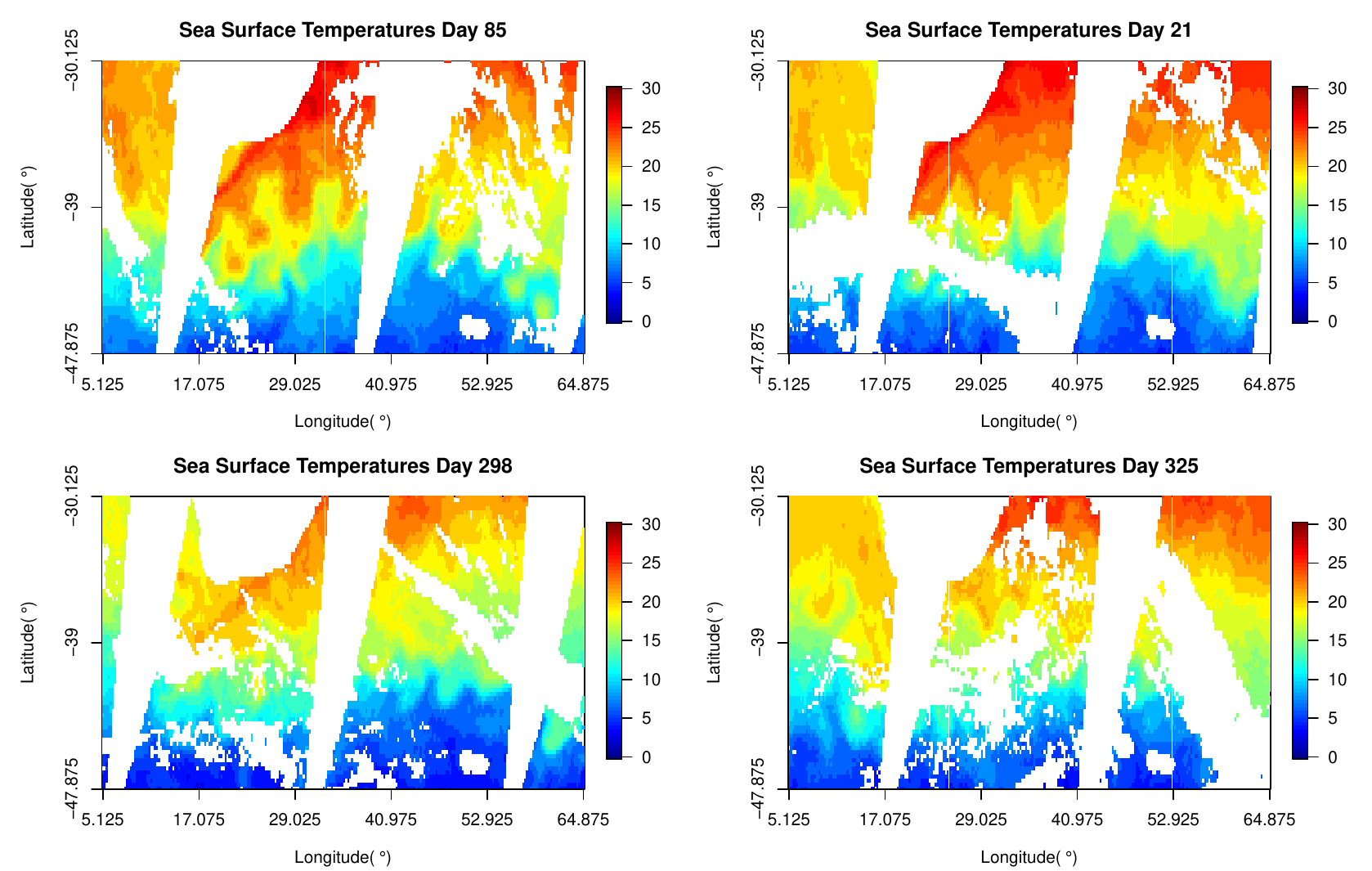}
 \caption{}
 \label{fig:SST_ori}
 \end{subfigure}
 \begin{subfigure}{0.8\textwidth}
 \includegraphics[width=0.99\linewidth]{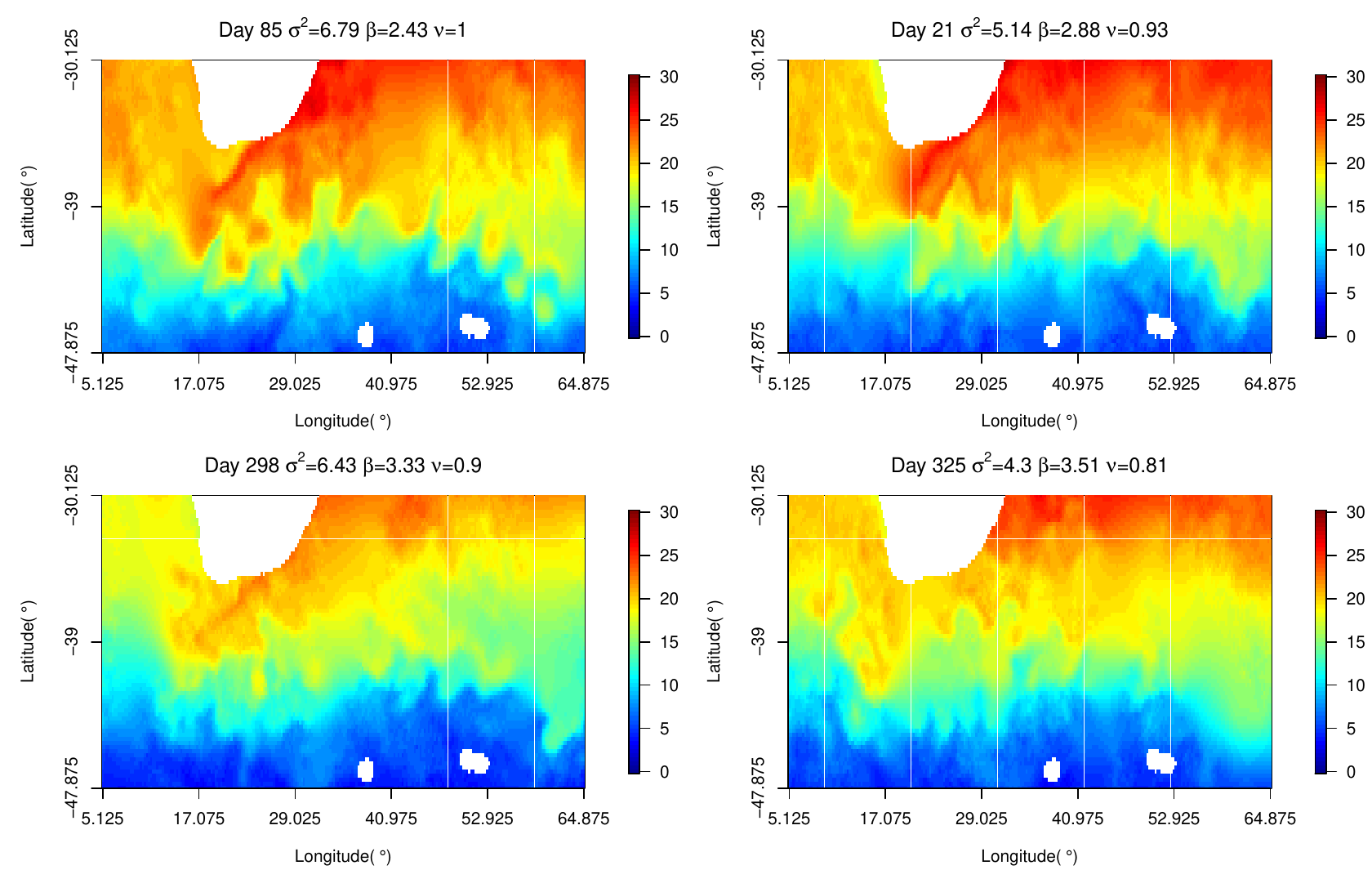}
 \caption{}
 \end{subfigure}
 \caption{Panels~(a) are the original sea surface temperatures in Celsius (\textdegree{}C) where the locations with NA values are colored in white. Panels~(b) are the predicted sea surface temperatures in Celsius (\textdegree{}C) based on the linear mean structure and the kriging results where the land area is not predicted and colored in white. Parameter estimates are provided for each panel.}
 \label{fig:pred_temp_cj}
\end{figure}

Our analysis considers only the spatial structure in the spatio-temporal
dataset and hence, assumes independence between the parameters on
different days. Before introducing our model, we first present some exploratory 
data analysis. In Figure~\ref{fig:SST_ori}, we use the {\em image.plot} 
function from the {\em fields} package to plot the heatmap of the sea surface 
temperature on selected four days that are also used for showing the kriging 
results later. Numerous gaps are present in the data, corresponding to 
three main causes: 1) land: specifically South Africa and Lesotho, visible 
in the left-center of the top of the plot, as well as two small islands towards 
the southern boundary; 2) clipping: the large wedge-shaped voids 
cutting N-S across the picture resulting from the satellite's orbital path; 
and 3) cloud cover: all or most of the remaining swirls and dots present in 
the image. Various forms of kriging can be used to attempt to fill those gaps 
caused by orbital clipping and cloud cover. Of course, it does not make sense 
to estimate sea surface temperatures for gaps caused by the presence of land. 
A pronounced temperature gradient is visible from highs of over 25$^\circ$ C in 
the north of the study area to a low of 3.5$^\circ$ C towards the southern 
boundary. This indicates spatial correlation in the dataset, but it also shows 
that the data are not stationary, as the mean temperature must vary considerably 
with latitude.

\begin{figure}[b!]
 \centering
 \includegraphics[width=0.95\linewidth]{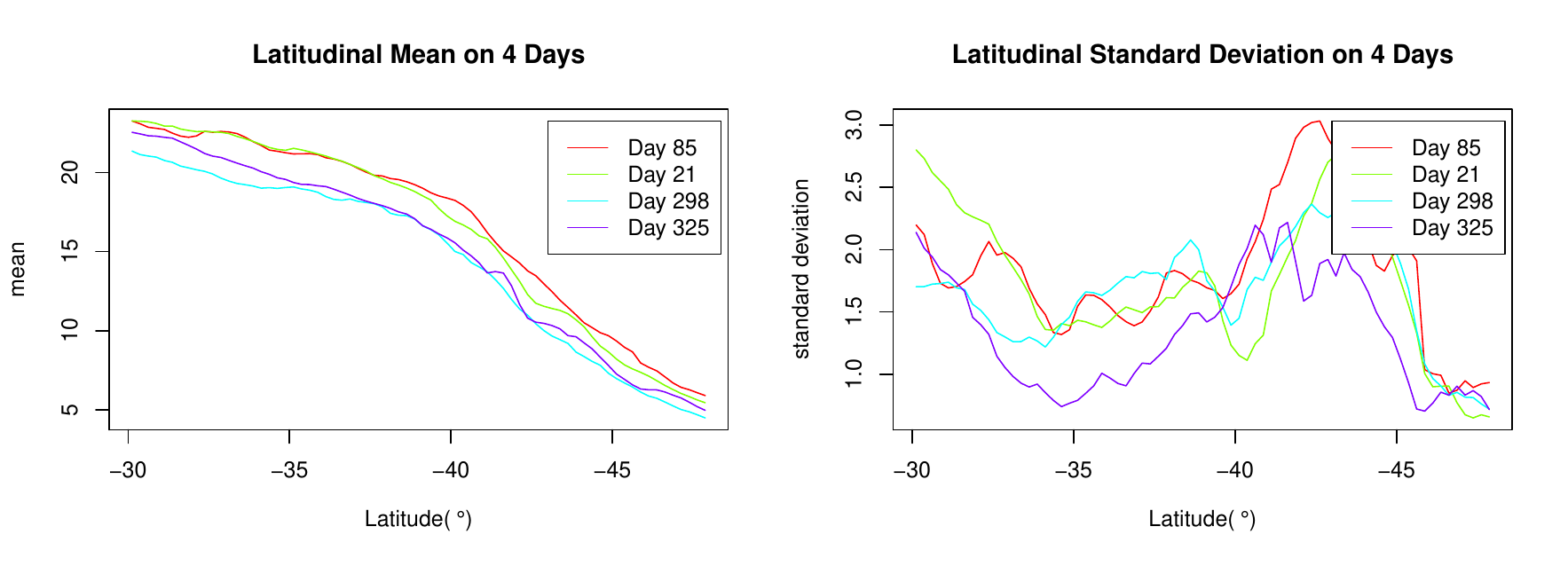}
 \caption{Exploratory data analysis on Days~85, 21, 298, and 325. The mean and standard deviation are computed with missing values removed.}
 \label{fig:EDA}
\end{figure}

We plot the mean and standard deviation along each latitude on the four days in Figure~\ref{fig:EDA}. 
The longitudinal mean and standard deviation (not shown) are relatively stable, although 
there are spikes and troughs due to the missing data. Since the locations are 
sufficient for a regression with only three parameters, we also include the longitude
as a regression variable and assume the following linear model with Gaussian noise 
for the sea surface temperature, $T(\lambda, \alpha)$:
\begin{align*}
T(\lambda , \alpha) &= \mu(\lambda , \alpha) + \epsilon(\lambda , \alpha ; \sigma^2 , \beta , \nu), \\
\mu(\lambda , \alpha) &= c + a \cdot \lambda + b \cdot \alpha, 
\end{align*}
where $\lambda$ denotes longitude, $\alpha$ denotes latitude, and $\epsilon(\lambda, \alpha)$ is a GRF with zero mean and a Mat{\'e}rn covariance structure parameterized as in (\ref{eq:matern}). Here, $a$, $b$, and $c$ are the linear coefficients for the mean 
structure, which we compute prior to the covariance structure based on the least 
square estimation. Maximum likelihood estimation is only applied to $ \sigma^2, \beta$, and $\nu$ at the second stage because non-convex optimization for six parameters requires a larger 
sample size and significantly more iterations. The original data have different proportions 
of missing values, varying from Day 1 to Day 331. We ignore those days whose missing 
value proportion exceeds $50\%$ so that the number of predictions is not more than the 
original number of observations.

The model fitting is done with the {\em {exact\_mle}} function from the \ExaGeoStatR package, which maximizes the exact likelihood. Specifically, the function call is:


\begin{lstlisting}[language=R, texcl=true,upquote=true,literate={--}{{-{}-}}1,   basicstyle=\ttfamily,basicstyle=\footnotesize,keywordstyle=\ttfamily, deletekeywords={ beta, data,frame,length,as,character},showstringspaces=false]
> x = x[!is.na(z)]
> y = y[!is.na(z)]
> z = z[!is.na(z)]
> mydf = data.frame(x, y, z)
> mymdl = lm(formula = z ~ x + y , data = mydf)
> z = as.numeric(mymdl$residuals)
> mytime = system.time(theta_out = exact_mle(data, kernel = "ugsm-s",
  dmetric = "euclidean", optimization = list(clb = c(0.001, 0.001, 0.001), 
  cub = c(5, 5, 5 ), tol = 1e-4, max_iters = 20))[3]
\end{lstlisting}

Referring to Section~\ref{section3}, {\em n} is the number of spatial 
locations, {\em ncore} and {\em ngpu} are the numbers of CPU cores 
and GPU accelerators to deploy, {\em ts} denotes the tile size used 
for parallelized matrix operations, {\em pgrid} and {\em qgrid} are the cluster
topology parameters, {\em x} and {\em y} store either the Cartesian 
coordinates or the spherical coordinates of the geometry, {\em z} is one 
realization of the spatial variables of dimension {\em n}, {\em clb} 
and {\em cub} define the search range for the three parameters, {\em dmetric} is a 
boolean indicating whether the Euclidean distance or the great circle 
distance is used for the covariance matrix, {\em tol} and {\em niter} specify 
the stopping criteria which supports both a tolerance level for reckoning 
convergence and the maximum number of iterations. This application 
study uses $16$ Intel Sandy Bridge Xeon E5-2650 processors without any GPU acceleration.

The tile size is initialized at $160$, and the grid dimensions are both $1$ for simplicity. 
In order to compare with the {\em geoR} and {\em fields} packages, we set {\em niter} 
to $20$ and measure the time cost of fitting the GRF to the data on Day~1, which has 
over $8{,}800$ valid (not NA) locations. The following is the code calling the {\em likfit} 
and {\em spatialProcess} functions while the function call for {\em {exact\_mle}} was 
already shown above:


\begin{lstlisting}[language=R, texcl=true,upquote=true,literate={--}{{-{}-}}1,   basicstyle=\ttfamily,basicstyle=\footnotesize,keywordstyle=\ttfamily, deletekeywords={ beta, data,frame,length,as,character},showstringspaces=false]
> time = system.time(fit_obj = spatialProcess(cbind(x, y), z, cov.args = list(
      Covariance = "Matern", smoothness = 0.8), verbose = T, theta.start = 0.1, 
      theta.range = c(0.1, 5), optim.args = list(method = "Nelder-Mead", 
      control = list(maxit = 20, tol = 1e-4))))[3]
> data_obj = as.geodata(cbind(x, y, z))
> time = system.time(fit_obj = likfit(geodata = data_obj, trend = "cte", 
      ini.cov.pars = c(0.1, 0.1), fix.nugget = TRUE, nugget = 0, fix.kappa = FALSE, 
      kappa = 0.1, cov.model = "matern", lik.method = "ML", limits = pars.limits(
      sigmasq = c(0.01, 20), phi = c(0.01, 20), kappa = c(0.01, 5)), print.pars = TRUE, 
      method = "Nelder-Mead", control = list(maxit = 20,abstol = 1e-4)))[3]
\end{lstlisting}

The {\em {exact\_mle}} function was executed with $15$ CPUs and 
took $147$ seconds, the {\em likfit} function from the {\em geoR} package 
cost $2{,}286$ seconds, and the {\em spatialProcess} function from 
the {\em fields} package needed $4{,}049$ seconds. It usually requires more 
than $100$ iterations to reach convergence, which renders the {\em geoR} 
and {\em fields} packages very difficult to use for fitting high-dimensional GRFs, whereas 
the \ExaGeoStatR package utilizes parallel architectures and reduces the time 
cost by more than one order of magnitude. Hence, \ExaGeoStatR allows to fill 
many spatial images quickly.

We select $(0.01 , 20.0)$ as the searching range for $\sigma^2$ and $\beta$ 
and $(0.01 , 5.00)$ for $\nu$ to guarantee the results not landing on boundary 
values. The initial values for all three parameters are the corresponding lower 
bounds of the searching ranges by default, and the optimization continues without 
any limit on the number of iterations until convergence is reached within a tolerance 
level of $10^{-4}$. There are $174$ days whose missing value percentages are 
below $50\%$, and Table~\ref{tbl:sum_sts_cj} summarizes the independently-estimated
parameters for these days.

Here, $\nu$ has the most consistent estimations among the three 
parameters while $\sigma^2$ and $\beta$ have similar variability. Based on 
the estimated parameters, we predict the sea surface temperature at locations 
where the data are missing with kriging, which computes the conditional 
expectation using a global neighborhood. The kriging is done with  {\em exact\_predict()} from the {\em ExaGeoStatR} package as indicated below:
\begin{lstlisting}[language=R, texcl=true,upquote=true,literate={--}{{-{}-}}1,   basicstyle=\ttfamily,basicstyle=\footnotesize,keywordstyle=\ttfamily, deletekeywords={ beta, data,frame,length,as,character},showstringspaces=false]
> hardware = list (ncores = 2, ngpus = 0, ts = 320, pgrid = 1, qgrid = 1)
> exageostat_init(hardware)
> Data_train_list <- list("x" = x_known, "y" = y_known, "z" = z_known )
> Data_predict_list <- list("x" = x_unknown, "y" = y_unknown)
> prediction_result <- exact_predict(Data_train_list, 
  Data_predict_list,"ugsm-s",  dmetric, est_theta, 0)
> exageostat_finalize()


\end{lstlisting}




In Figure~\ref{fig:pred_temp_cj} we show the original and the predicted sea
surface temperature for the four days corresponding to the $99\%$, $66\%$, $33\%$, and $1\%$ quantiles of the estimated $\nu$ to visualize the smoothness 
change. Day~85 seemingly has more details than Day~325, although 
the main factor governing the temperature change is the mean structure.

\section{Discussion}
\label{section5}

Statistical modeling methods have been widely used to analyze and 
understand the behavior of geospatial data in environmental data science applications. 
For example, the Maximum Likelihood Estimation (MLE) is used to model 
environmental data by building a covariance matrix to describe the relations 
between the observations at geographical locations. This operation has $O(n^3)$ 
computation complexity and $O(n^2)$ memory complexity
due to the need to perform an inverse function to a generated covariance matrix 
with dimension equal to the number of locations, $n$. 

Environmental data have increased tremendously in size
due to recent advances in observational techniques and existing tools cannot
easily handle them, especially in the statisticians' preferred programming
environment, i.e., \R. Therefore, in this work, we tackled the computational
complexity of the MLE operation on large-scale systems within the \R environment. We
presented the \ExaGeoStatR package that allows large-scale geospatial modeling and
prediction using the MLE operation. By exploiting the current state-of-the-art parallel 
linear algebra solvers and dynamic runtime systems, \ExaGeoStatR can compute
the Gaussian maximum likelihood function on different parallel 
architectures (i.e.,  shared memory, GPUs, and distributed systems). Large-scale
Gaussian calculations in \R become possible with \ExaGeoStatR by 
mitigating its memory space and computing restrictions.

The \ExaGeoStatR package provides four options to evaluate the MLE 
operation, i.e., exact, Diagonal Super-Tile (DST), Tile Low-Rank (TLR), 
and Mixed-Precision (MP). However, this work was only concerned with 
analyzing and assessing the performance of the exact computation variant (the focus of comparison in this paper) against existing well-known \R packages, such as \geor and \fields. We focused on exact computations to highlight the parallel capabilities of  \ExaGeoStatR with the exact solution over the aforementioned \R packages and its ability to run on different hardware architectures. Of course, parallelizing the approximation methods and scaling them on a large system will allow crossing the memory threshold of these methods with large problem sizes. However, we did not cover the approximation capabilities of \ExaGeoStatR since it was well-covered in previous studies~\citep{abdulah2018exageostat, abdulah2018parallel,abdulah2019geostatistical,hong2021efficiency,abdulah2021accelerating}.

The evaluation demonstrates a significant 
difference in \ExaGeoStatR performance compared to the other 
two packages. The accuracy evaluation also shows that \ExaGeoStatR 
performs very well on synthetic datasets compared to \geor and  \fields. The 
evaluations of the other \ExaGeoStatR computation methods (DST, TLR, MP) can be found in \cite{abdulah2018exageostat, abdulah2018parallel, abdulah2019geostatistical,abdulah2021accelerating}.  
\ExaGeoStatR  also includes seven covariance functions.
In this work,
we evaluated the performance of univariate stationary Gaussian random fields with Mat{\'e}rn covariance function. Other 
studies include the evaluation of other covariance functions such as multivariate models~\citep{salvana2021high}, and space-time models~\citep{10.1145/3539781.3539800}. \exageostatr can also include a nugget effect in all of its kernels.

We aimed from the beginning to abstract the parallel execution functions
away from the concerns of the \R developer. 
The developer only needs to specify some parameters to define the underlying
hardware architecture and the package will automatically optimize the execution on
the target hardware through the internal runtime system. In this way, we improve the
portability of our software and make it more suitable for the \R community developers.
The current version of \ExaGeoStatR only supports a zero mean to provide a robust and efficient estimation of covariance. However, the package can also be helpful in many other problems. First, when
the mean function is not zero, the simplest approach is to estimate the mean and
the covariance function independently, as we did in the application tutorial. Theoretically, this
independent maximum likelihood estimation will result in a biased random effect and
can be improved by the restricted maximum likelihood (REML) techniques. 
However, as {\fields} suggests, using REML typically does not make much 
difference. Second, we can predict at new locations (kriging) with uncertainties 
once the covariance parameters are estimated. The prediction is calculated by 
the conditional distribution of the multivariate Gaussian. Third, even when spatial 
nonstationarity is observed, we can still apply \ExaGeoStatR by assuming local 
stationarity. This idea is implemented in {\em convoSPAT} \citep{Risser:2017aa}. Once we 
obtain the estimated parameters locally, we can reconstruct the nonstationary 
covariance function. Finally, \ExaGeoStatR is also useful for space-time and 
multivariate GRFs, where the covariance function we use should be replaced 
by a spatio-temporal covariance function and a cross-covariance function, respectively. 
As for our future work, \ExaGeoStatR will provide the necessary built-in functions 
to support the extensions mentioned above for more complex applications.

\section{Acknowledgment}
The research in this manuscript was supported by
funding from the King Abdullah University of Science and
Technology (KAUST) in Thuwal, Saudi Arabia. We would like to thank the Supercomputing Laboratory (KSL) at KAUST for providing the hardware resources used in this work, including the Ibex cluster and Shaheen-II Cray XC40 supercomputer. Finally, the authors would
like to thank Bilel Hadri and Greg Wickham from the KSL team for their valuable help in running the experiments in this publication.



%
\baselineskip 21 pt
\bibliographystyle{abbrvnat}
\bibliography{exageostatr}

%
\newpage
\section*{Appendix: ExaGeoStatR Installation Tutorial}
\begin{appendices}
Herein, we provide a detailed description of how to install the \ExaGeoStatR package with different capabilities.  \ExaGeoStatR is currently supported on both macOS  and Linux systems. To automatically compile several dependencies, \ExaGeoStatR requires a source of BLAS, CBLAS, LAPACK and LAPACKE routines
(e.g., Intel MKL and OpenBLAS) that must be available on the system before
installing \ExaGeoStatR. The package is hosted in a GitHub repository and can be downloaded and installed directly from there. The \ExaGeoStatR package is available 
through \href{https://github.com/ecrc/exageostatR}{https://github.com/ecrc/exageostatR} GitHub repository. We also provide an \ExaGeoStatR docker image to increase the reusability of the package. The docker image can be found at  \href{https://hub.docker.com/r/ecrc/exageostat-r}{https://hub.docker.com/r/ecrc/exageostat-r}.

\ExaGeoStatR includes a self-installation configuration file that helps the user installation of different software dependencies as mentioned in Section II. Thus, to directly install \ExaGeoStatR from GitHub:


\begin{lstlisting}[language=R, texcl=true,upquote=true,literate={--}{{-{}-}}1,   basicstyle=\ttfamily,basicstyle=\footnotesize,keywordstyle=\ttfamily, deletekeywords={ beta, data,frame,length,as,character},showstringspaces=false]
> install.packages("devtools")
> library("devtools")
> install_git(url = "https://github.com/ecrc/exageostatR")
\end{lstlisting}

The $install\_git$ command can be edited to change the default configuration of the \ExaGeoStatR package to support
several installation modes:


\begin{itemize}

\item To enable \MPI support for distributed memory systems (i.e., an \MPI library should be available on the system (e.g. MPICH, OpenMPI, and IntelMPI)):

\begin{lstlisting}[language=R, texcl=true,upquote=true,literate={--}{{-{}-}}1,   basicstyle=\ttfamily,basicstyle=\footnotesize,keywordstyle=\ttfamily, deletekeywords={ beta, data,frame,length,as,character},showstringspaces=false]
> install_git(url = "https://github.com/ecrc/exageostatR", 
  configure.args = c('--enable-mpi'))
\end{lstlisting}


\item To enable CUDA support for GPU systems (i.e., the CUDA library should
be available on the system):
\begin{lstlisting}[language=R, texcl=true,upquote=true,literate={--}{{-{}-}}1,   basicstyle=\ttfamily,basicstyle=\footnotesize,keywordstyle=\ttfamily, deletekeywords={ beta, data,frame,length,as,character},showstringspaces=false]
> install_git(url = "https://github.com/ecrc/exageostatR", 
  configure.args = c('--enable-cuda'))
\end{lstlisting}


\item If all \ExaGeoStatR software dependencies have been already installed
on the system  (i.e., install \ExaGeoStatR package without
dependencies):

\begin{lstlisting}[language=R, texcl=true,upquote=true,literate={--}{{-{}-}}1,   basicstyle=\ttfamily,basicstyle=\footnotesize,keywordstyle=\ttfamily, deletekeywords={ beta, data,frame,length,as,character},showstringspaces=false]
> install_git(url = "https://github.com/ecrc/exageostatR", 
  configure.args = c('--no-build-deps'))
\end{lstlisting}

\end{itemize}

The Docker image can be also an easy way to use the \ExaGeoStatR package but the performance could be impacted on different hardware architectures. The Docker pull command for \ExaGeoStatR package is:
\begin{lstlisting}[language=bash,
texcl=true,upquote=true,literate={--}{{-{}-}}1,   basicstyle=\ttfamily,basicstyle=\footnotesize,keywordstyle=\ttfamily, deletekeywords={ beta, data,frame,length,as,character},showstringspaces=false]
docker pull ecrc/exageostat-r
\end{lstlisting}

To independently install {\em ExaGeoStatR} dependencies, the user can follow the complete installation guide at: \href{https://github.com/ecrc/exageostatR/blob/master/InstallationGuide.md}{https://github.com/ecrc/exageostatR/blob/master/InstallationGuide.md}

\begin{table}[!p]
\centering
\caption{Comparison of some existing Gaussian process software.}
\label{tab:pkgs}
\resizebox{0.8\textwidth}{!}{%
\begin{tabular}{|c|c|c|c|c|c|c|}
\hline
Package & Platform & Version &\begin{tabular}[c]{@{}c@{}}Exact\\ Calc.\end{tabular} & \begin{tabular}[c]{@{}c@{}}Approx.\\ Calc.\end{tabular} & \begin{tabular}[c]{@{}c@{}}Supports Parallel \\ Execution\end{tabular}& Reference \\ \hline
\biggp & R &V0.1-7& \cmark  & \xmark  & \cmark  &  \citep{bigGP}\\ \hline
\exageostatr & R & V1.0.1 & \cmark   & \begin{tabular}[c]{@{}c@{}}\cmark \end{tabular} & \cmark &  This work\\ \hline
\fields & R & V14.1 & \cmark  & \xmark   & \xmark&  \citep{fields}\\ \hline
\geor & R & V1.9-2 & \cmark & \xmark  & \xmark& \citep{geoR} \\ \hline
{\em GPfit} & R & V1.0-8 & \cmark  & \xmark  &\xmark &  \citep{macdonald2015gpfit}\\ \hline
{\em GpGp} & R & V0.4.0 & \xmark & \cmark   & \xmark& \citep{guinness2021gaussian} \\ \hline
{\em GPvecchia}& R & V0.1.3 & \cmark  & \cmark  & \xmark  & \citep{katzfuss2020gpvecchia}\\ \hline
{\em GPy} & Python & V1.0.7 & \cmark  & \cmark   &  \xmark& \citep{matthews2017gpflow} \\ \hline
{\em gstat}& R & V2.0-9 & \cmark  & \xmark  & \xmark  & \citep{pebesma2004multivariable}\\ \hline
 \inla & R & V22.05.07 &\xmark  & \cmark   & \cmark&  \citep{lindgren2015bayesian}\\ \hline
 {\em LaGP} & R & V1.5-7 & \xmark & \begin{tabular}[c]{@{}c@{}}\cmark \end{tabular}  & \cmark  & \citep{gramacy2016lagp} \\ \hline
 
 {\em mlegp}& R & V3.1.9 & \cmark  & \xmark   &  \begin{tabular}[c]{@{}c@{}}\cmark \end{tabular}&  \citep{dancik2008mlegp}\\ \hline
 \randomfields & R & V3.1.50 & \cmark  & \xmark  & \xmark& \citep{schlather2015analysis} \\ \hline

\spbayes & R & V0.4-6 & \cmark  & \xmark  & 
\begin{tabular}[c]{@{}c@{}}\cmark \end{tabular} &
\citep{finley2007spbayes}\\ \hline

\end{tabular}}
\end{table}

\begin{table}[!p]
\caption{\label{tab:software_dep} \ExaGeoStat software dependencies.}
\centering
\resizebox{0.8\textwidth}{!}{%
\begin{tabular}{lp{10cm}}
\hline
Software           & Description \\ \hline

\nlopt  & Nonlinear optimization library: provides a common interface for several optimization algorithms implementations.\\ \hline

\chameleon& A dense linear algebra software relying on sequential task-based algorithms and dynamic runtime systems. \\ \hline

\hicma & Hierarchical Computations on Manycore Architectures: a low rank matrix computation library exploiting the data sparsity of the matrix operator. \\ \hline

\dplasma & A dense linear algebra package for distributed heterogeneous systems. \\ \hline
 
\starpu &A runtime system library for task-based programming model running on shared/distributed-system architectures as well as GPU-based systems.\\ \hline

\parsec &A generic framework for architecture aware scheduling and management of micro-tasks on distributed many-core heterogeneous architectures.\\ \hline

\starsh & Software for Testing Accuracy, Reliability, and Scalability of Hierarchical computations: a high performance low-rank matrix approximation library generating low-rank matrices on shared/distributed-memory systems. \\ \hline

\mkl/\openblas &  Optimized linear algebra libraries implementations for CPU/GPU. \\ \hline

 \hwloc& Portable Hardware Locality: provides a portable abstraction of the hierarchical topology of modern architecture.\\ \hline
 
\gsl& {\em GNU} Scientific Library: provides a set of numerical computing routines.\\ \hline

\end{tabular}
}
\end{table}

\begin{table}[!p]
\caption{\label{tab:egsr_funs} Overview of \ExaGeoStatR functions.}
\centering
\resizebox{0.8\textwidth}{!}{%
\begin{tabular}{lp{9cm}}
\hline
Function Name           & Description \\ \hline

{\em exageostat\_init}  &         Initiate \ExaGeoStat instance, defining the underlying hardware (i.e., number of CPU/GPU cores  and the tile size). \\ \hline
{\em simulate\_data\_exact} &             Generate $\mathbf{z}$ measurements vector at $n$ unstructured random 2D locations. \\ \hline
{\em simulate\_obs\_exact} &              Generate $\mathbf{z}$ measurements vector at $n$ given 2D locations. \\ \hline
{\em exact\_mle} &              Compute the MLE model parameters (exact computation). \\ \hline
{\em dst\_mle}&               Compute the MLE model parameters (DST approximation computation). \\ \hline
{\em tlr\_mle} &              Compute the MLE model parameters (TLR approximation computation). \\ \hline
{\em mp\_mle} &              Compute the MLE model parameters (mixed-precision approximation computation). \\ \hline

{\em exact\_predict} &              Predict measurements at new locations with given model parameters (exact computation). \\ \hline

{\em exact\_mloe\_mmom} &              Compute MLOE and MMOM metrics~\citep{hong2021efficiency} based on new locations with given model parameters (exact computation). \\ \hline

{\em exact\_fisher} &              Compute Fisher information matrix with given model parameters (exact computation). \\ \hline
{\em exageostat\_finalize}&              Finalize current active \ExaGeoStat instance. \\ \hline

\end{tabular}
}
\end{table}

\begin{table}[!p]
\caption{\label{tab:software_kernels} \ExaGeoStatR supported covariance functions.}
\centering
\resizebox{0.8\textwidth}{!}{%
\begin{tabular}{lp{10cm}}
\hline
Kernel          & Description \\ \hline

{\em ugsm-s}  & Univariate Gaussian stationary Mat{\'e}rn - space \\ \hline


{\em bgsfm-s} & Bivariate Gaussian stationary flexible Mat{\'e}rn - space  \\ \hline

{\em bgspm-s} & Bivariate Gaussian stationary parsimonious Mat{\'e}rn - space \\ \hline

{\em tgspm-s} & Trivariate Gaussian stationary parsimonious Mat{\'e}rn - space \\ \hline
 
{\em ugsm-st} & Univariate Gaussian stationary Mat{\'e}rn - space-time\\ \hline

{\em bgsm-st} & Bivariate Gaussian stationary Mat{\'e}rn - space-time\\ \hline

\end{tabular}
}
\end{table}

\begin{table}[!p]
\centering
\caption{Differences between the estimation functions of {\geor }, {\fields }, and  \exageostatr}
\resizebox{0.9\textwidth}{!}{%
\begin{tabular}{c|c|c|c}
\hline
\hline
Package & {\geor} & {\fields} & \exageostatr\\
\hline
Function name  &{\em likfit} & {\em spatialProcess} &{\em exact\_mle}\\
Mean & estimated & estimated & fixed as zero\\
Variance & estimated & estimated & estimated\\
Spatial Range & estimated & estimated & estimated\\
Smoothness  & estimated & fixed & estimated\\
Default optimization method & {\em Nelder-Mead} & {\em BFGS}$^{1}$& {\em BOBYQA}$^{2}$\\
\hline
\hline
\multicolumn{4}{l}{\textsuperscript{1}\footnotesize{BFGS: Broyden-Fletcher-Goldfarb-Shanno \textsuperscript{2}BOBYQA: bound optimization by quadratic approximation}}

\end{tabular}
}
\label{tab:packages}
\end{table}

\begin{table}[!p]
\centering
\caption{The average execution time per iteration  and the average number of iterations to reach the tolerance based on 100 samples. Nine scenarios with three smoothness parameters, $\nu$, and three spatial ranges, $\beta$, are assessed. The variance, $\sigma^2$, is set to be one. Smallest values are in bold.}
\resizebox{0.70\textwidth}{!}{%
\begin{tabular}{c|ccc|ccc|ccc}
\hline
\hline
\multicolumn{10}{c}{The average execution time per iteration (seconds)}\\
\hline
Package & \multicolumn{3}{c}{{\geor}}&\multicolumn{3}{c}{{\fields}} &  \multicolumn{3}{c}{\exageostatr}\\
\hline
\backslashbox{$\nu$=}{$\beta$=} & 0.03&0.1& 0.3& 0.03&0.1& 0.3& 0.03&0.1& 0.3\\
\hline
$0.5$ & 1.39&1.49&1.47&0.75&0.97&0.99& \textbf{0.10} & \textbf{0.12} & \textbf{0.12}\\
$1$ & 1.35&1.49&1.56& 0.66&0.90&0.90 & \textbf{0.09} & \textbf{0.13} & \textbf{0.13}\\
$2$ &  1.34&1.56&1.57&0.67&0.91&0.93 & \textbf{0.09} & \textbf{0.13} & \textbf{0.13}\\
\hline
\hline
\multicolumn{10}{c}{The average number of iterations to reach the tolerance}\\
\hline
\backslashbox{$\nu$=}{$\beta$=} & 0.03&0.1& 0.3& 0.03&0.1& 0.3& 0.03&0.1& 0.3\\
\hline
$0.5$ & 160&157&135&\textbf{73}&\textbf{72}&\textbf{70}& 231&204&237\\
$1$ &  193&\textbf{33}&\textbf{23}&\textbf{75}&{75}&{80} & 318&320&275\\
$2$  & 216&\textbf{25}&\textbf{20}& \textbf{100}&70&{85}& 427&436&332\\
\hline
\hline
\end{tabular}}
\label{tab:iters}
\end{table}

\begin{table}[!p]
 \centering
 \caption{Summary statistics for the estimated parameters across 174 days of sea surface temperature.}
 \label{tbl:sum_sts_cj}
 \resizebox{0.65\textwidth}{!}{%
 \begin{tabular}{l|r|r|r|r|r|r}
   \hline
  \hline
  & Min & 25\% Q & Median & Mean & 75\% Q & Max \\
  \hline
  $\sigma^2$ & $3.41$ & $5.78$ & $6.44$ & $6.33$ & $6.76$ & $14.40$ \\
  $\beta$ & $1.99$ & $2.76$ & $3.02$ & $3.03$ & $3.27$ & $4.60$ \\
  $\nu$ & $0.81$ & $0.89$ & $0.91$ & $0.91$ & $0.93$ & $1.00$\\
  \hline 
  \hline
 \end{tabular}}
\end{table}

\end{appendices}




\end{spacing}
\end{document}